\documentclass[5p, twocolumn]{elsarticle}

\makeatletter
\def\ps@pprintTitle{%
 \let\@oddhead\@empty
 \let\@evenhead\@empty
 \def\@oddfoot{}%
 \let\@evenfoot\@oddfoot}
\makeatother

\usepackage{graphicx}
\usepackage{caption}
\usepackage{subcaption}
\graphicspath{{./figures/}{./}}

\usepackage{amssymb}
\usepackage{amsthm}
\usepackage{amsmath}

\usepackage{lineno}
\usepackage{hyperref}
\hypersetup{
    colorlinks=true,
    linkcolor=blue,
    filecolor=magenta,      
    urlcolor=cyan,
}
\modulolinenumbers[5]
\usepackage{layouts}
\newcommand{\figsize}{9.17cm}

\newcommand*{\myeqref}[2][Eq.~]{%
  \hyperref[{#2}]{#1(\ref*{#2})}%
}

\usepackage{numcompress}

\begin{document}

\begin{frontmatter}



\title{Temporal Markov Processes for Transport in Porous Media: Random Lattice Networks}


\author[farm]{Amir H.~Delgoshaie \corref{cor}}
\ead{amirdel@stanford.edu}
\author[eth]{Patrick Jenny}
\ead{jenny@ifd.mavt.ethz.ch}
\author[farm]{Hamdi A.~Tchelepi}
\ead{tchelepi@stanford.edu}
\cortext[cor]{Corresponding author}
\address[farm]{Department of Energy Resources
Engineering, Stanford University, Stanford, CA, USA}
\address[eth]{Institute of Fluid Dynamics, ETH Z\"urich, Z\"urich, Switzerland}

\begin{abstract}
Monte Carlo (MC) simulations of transport in random porous networks indicate that for high variances
of the log-normal permeability distribution, the transport of a passive tracer is non-Fickian. Here
we model this non-Fickian dispersion in random porous networks using discrete temporal Markov
models.  We show that such temporal models capture the spreading behavior accurately. This is true
despite the fact that the slow velocities are strongly correlated in time, and some studies have
suggested that the persistence of low velocities would render the temporal Markovian model
inapplicable. Compared to previously proposed temporal stochastic differential equations with
case-specific drift and diffusion terms, the models presented here require fewer modeling
assumptions.  Moreover, we show that discrete temporal Markov models can be used to represent
dispersion in unstructured networks, which are widely used to model porous media. A new method is
proposed to extend the state space of temporal Markov models to improve the model predictions in the
presence of extremely low velocities in particle trajectories and extend the applicability of the
model to higher temporal resolutions.  Finally, it is shown that by combining multiple transitions,
temporal models are more efficient for computing particle evolution compared to correlated CTRW with
spatial increments that are equal to the lengths of the links in the network.  \end{abstract}

\end{frontmatter}


\section{Introduction}
Modeling transport in porous media is highly important in various applications including water
resources management and extraction of fossil fuels. Predicting flow and transport in aquifers and
reservoirs plays an important role in managing these resources. A significant factor influencing
transport is the heterogeneity of the flow field, which results from the underlying heterogeneity of
the conductivity field. Transport in such heterogeneous domains displays non-Fickian characteristics
such as long tails for the first arrival time probability density function (PDF) and non-Gaussian
spatial distributions \citep{berkowitz2006modeling, bouchaud1990anomalous, edery2014origins}.
Capturing this non-Fickian behavior is particularly important for predictions of contaminant
transport in water resources. For example, in water resources management long tails of the arrival
time PDF can have a major impact on the contamination of drinking water, and therefore efficient
predictions of the spatial extents of contaminant plumes is key
\citep{nowak2012hypothesis,moslehi2017uncertainty,ghorbanidehno2015real, li2015compressed,
ghorbanidehno2017optimal}.\\

Past studies have provided a range of models for predicting this non-Fickian transport. The
continuous time random walk (CTRW) formalism offers a framework to study anomalous transport through
disordered media and networks \citep{berkowitz2006modeling, fiori2015advective}.  However, in most
studies where this approach is used, velocity correlation between successive tracer particle jumps
are neglected. The time domain random walk method (TDRW), which is conceptually similar to the CTRW
method, directly calculates the arrival time of a particle cloud at a given location
\citep{banton1997new, bodin2003simulation}. Similar to CTRW, consecutive velocities resulting from
the TDRW method are independent of each other. Detailed studies of transport have  shown
conclusively that particle velocities in mass conservative flow fields are correlated
\citep{WRCR:WRCR4722, le2007characterization}. To account for this correlation, Markov velocity
models have been developed. These models can be divided into three main groups of temporal, spatial,
and mixed (temporal and spatial) models based on the variables chosen to index the stochastic
velocity process.\\ 

\citet{le2008lagrangian} proposed discrete Markov chains for modeling the velocity process and
tested the Markov assumption for the longitudinal component of the velocity of tracer particles in
heterogeneous porous media. They studied transition probabilities for the velocity process in time
and space and concluded that spatial Markov models can characterize the velocity field, but in their
study temporal models were found to be unfit for this task. A one-dimensional spatial Markov model
was then used in \citep{le2008spatial} to successfully model transport in heterogeneous domains.
\citet{kang2011spatial, kang2015anomalous, kang2015impact} extended the spatial Markov model
framework to two dimensions and performed several studies on random lattice networks.  The spatial
Markov model was also applied to the velocity field resulting from simulation of flow in the pore
space of real rock and to disordered fracture networks \citep{kang2014,kang2017anomalous}.  Meyer et
al. used a temporal Markov model and successfully modeled particle dispersion in two-dimensional
cases using stochastic differential equations \citep{meyer2010particle}. This framework was then
used to model the joint velocity-concentration PDF \citep{meyer2010joint}. In another study,
\citet{meyer2016testing} provided a framework for testing the Markov hypothesis for the velocity of
tracer particles. Mixed temporal and spatial models have also been proposed.  \citet{meyer2013fast}
proposed a set of SDEs for modeling transport in exponential permeability fields with a velocity
process in time and an angle process in space. More recently, another mixed set of SDEs were
proposed to model the velocity process resulting from direct numerical simulation (DNS) of flow and
transport in the pore-space of real rocks \citep{meyer2016pore}.\\

Although temporal Markov velocity models have been shown to perform well in several studies, four
important gaps remain in the literature regarding the validity and potential of temporal Markov
velocity models.  First, the temporal Markov models that have been successfully applied for modeling
transport in porous media are stochastic differential equations (SDEs) with specific drift and
diffusion terms.  The drift and diffusion functions vary for different studies and are constructed
for the specific problem in each study \citep [e.g. compare][]{meyer2010particle, meyer2016pore}. In
this work we use discrete Markov chains which do not require modeling the functional form of the
drift and diffusion terms.\\

Second, we apply temporal Markov chains to model networks with perfect mixing at the nodes which
covers an important gap in the available literature. In two dimensions, particle tracking through a
network is fundamentally different from particle tracking in continuum scale (whether the continuum
scale is the pore-space of a rock or a permeability field). All the previous works on temporal
velocity models have investigated transport in porous media at the continuum scale, where different
streamlines cannot cross. However, in network models which are the subject of this paper,
streamlines do cross. At the continuum scale, the ensemble plume in two-dimensional domains will
asymptotically stop spreading in the transverse direction \citep{attinger2004exact}; however, the
second moment of the ensemble plume in networks do not necessarily reach an asymptote
\citep{kang2011spatial}. A temporal Markov model designed for two-dimensional continuum scale
problems (such as \citep{meyer2010particle}) cannot be applied to networks without modification.  \\
Third, the networks used to describe realistic porous media are in many cases unstructured
\citep{blunt2002detailed, dong2009pore, khayrat2016subphase, mehmani2017minimum} and having simple
efficient transport models for these networks is of great value. 
In addition to network models, recent developments in non-local models for flow in porous media
also call for simple transport models that are general enough for cases with distributions of
link lengths and transmissibilities \citep{delgoshaie2015non, jenny2016non}. 
Here the proposed temporal models are applied to both structured and unstructured networks. \\
Finally, a main argument against temporal Markov models with small average time steps is their
inability to capture the slow portions of the particle trajectories
\citep{le2008lagrangian,meyer2016testing}. In this work we show that temporal Markov models with
small time steps can be improved by adding information about the number of repetitions for a
velocity vector to their state definition. \\

In summary, here we study the performance of discrete temporal Markov models on random lattice
networks.  These networks are chosen to compare the performance of temporal Markov models with
existing correlated spatial models \citep{kang2011spatial}.  Temporal Markov models are used to
model transport in both structured and unstructured networks, and we show that these models can
yield accurate results in both cases. In contrast to spatial Markov models, the temporal models
proposed here do not require any modification for simulating transport in unstructured networks.
Moreover, we show that by combining multiple velocity transitions, temporal Markov models can be
more efficient compared with correlated CTRW with a jump extent equal to the network link length
\citep{kang2011spatial}.  Compared to temporal SDEs with specific drift and diffusion functions
\citep{meyer2010particle, meyer2013fast}, the models presented here contain significantly fewer
modeling assumptions, which makes them easier to apply to new problems (e.g. with a different
permeability correlation structure).  Finally, a novel way of enriching the state space for temporal
Markov models is discussed to reduce errors by accounting for velocity persistence when modeling
very slow transitions. The range of applicability of discrete temporal Markov models is studied by
quantifying the model prediction error for the spatial distribution of the particle plume and first
passage time distributions.

\section{The single-phase transport problem}
In both pore- and Darcy-scale problems, networks can be used for
transport modeling.  A network is defined by a set of nodes and a set of links connecting these
nodes.  
The transmissibility of the links, determines the strength of the
connection between two nodes. One can assume a linear relationship similar to Darcy's law for the
fluid flux $u_{ij}$ between the nodes $i$ and $j$, $u_{ij} = -k_{ij}(\Phi_j - \Phi_i)/l_{ij}$, where
$\Phi_i$ and $\Phi_j$ are the flow potentials, $k_{ij}$ is the connectivity of the link between the
two nodes and $l_{ij}$ is the length of the link. By defining $\gamma_{ij} = k_{ij}/l_{ij}$ as the
transmissibility of the link, the flux is the product of the link transmissibility and the potential
difference between the two nodes.\\


Once the fluxes in the links are known, one can simulate the transport of a passive tracer by
particle tracking. At each node the particle randomly chooses a link carrying flux out of that node
with a probability proportional to the flux in that link and travels along the selected link until
it arrives at a new node. \\
The movement of the particles can be modeled by a set of Langevin equations describing the particle
movement in space and time. In most studies, this set of Langevin equations has been stated in
either of two different ways, and these different prospects have led to Markov models in time and in
space for the particle
velocity.\\

In general, one can consider the velocity of a particle  after it has moved a distance
$\Delta s$, which may or may not be equal to the network link length or the grid spacing. On a
structured network, this would correspond to saving the velocity after each
transition to a new link.  We refer to the resulting velocity process as a spatial velocity process.
The resulting Langevin equations are

\begin{equation}
\begin{gathered}
\delta t_n = \Delta s / |\mathbf{v}_n|,\\
t_n = t_{n-1} + \delta t_n\\
\text{and } \mathbf{x}_n = \mathbf{x}_{n-1} + \mathbf{v}_n\delta t_n.
\end{gathered}
\label{eqSpace}
\end{equation}
Here, the subscript $n$ refers to the $n$th displacement of length $\Delta s$. Knowing all the
previous velocity vectors, the location of a particle can be found by using
Equations~\eqref{eqSpace}.  In published studies on applications of spatial Markovian models on
structured networks, 
spatial velocity processes are usually defined by consecutive displacements of a particle to new
nodes.  
One can also consider the average velocity of a particle in consecutive time steps of size $\Delta
t$. This
is referred to as a temporal velocity process. The Langevin equations for a temporal velocity
process are 
\begin{equation}
\begin{gathered}
t_n = t_{n-1} + \Delta t\\
\text{and } \mathbf{x_n} = \mathbf{x}_{n-1} + \mathbf{v}_n\Delta t.
\end{gathered}
\label{eqTime}
\end{equation}
Here, $n$ refers to the end of the $n$th time step. In the next section, we discuss the Markov
processes that have been introduced to efficiently
integrate the Langevin eqautions~\eqref{eqSpace} and \eqref{eqTime}.

\section{The Markov hypothesis}
In order to efficiently integrate the Langevin equations \eqref{eqSpace} and  \eqref{eqTime},
Markovianity is assumed for the velocity of the particles. The Markov hypothesis for the stochastic
velocity 
process, $\{\mathbf{v}_n, \;n=1,2,\dots\}$, assumes that the next state of the particle
velocity depends on the current state and is statistically independent of the rest of the history
of the process. That is, 

\begin{equation}
p(\mathbf{v}_n|\mathbf{v}_{n-1}, \mathbf{v}_{n-2}, \dots, \mathbf{v}_0) = 
p(\mathbf{v}_n|\mathbf{v}_{n-1}).
\end{equation}
To model transport using the Markov hypothesis, the transition
probabilities between different velocity states, $p(\mathbf{v}_n|\mathbf{v}_{n-1})$, needs to be
estimated. 
In two dimensions, the particle velocity is a vector in $\mathbb{R}^2$, and the state space for 
a Markov model
would be infinite.\\
One approach to model this process is to use discrete bins for the velocity vector and find a
discrete transition matrix between these bins. This is the approach taken by \citet{kang2011spatial}
for spatial Markov models and  \citet{le2008lagrangian, le2008spatial} for both temporal and spatial
Markov models.
An alternative approach, explored by \citet{meyer2010particle}, is to use SDEs to describe the
velocity process.
One important difference between the studies
by Le Borgne et al. and Meyer et al. is that in the latter works $p(\mathbf{v}_n|\mathbf{v}_{n-1})$
is not directly modeled, and the process is characterized using analytic drift and diffusion
functions for an SDE. An advantage of using SDEs is that explicit binning is not required for the
velocity classes. On the other
hand, more insight is needed to choose functions that can properly characterize the velocity
process. Functional representations have also been used in the context of spatial Markov models
\citep{kang2016emergence, painter2005upscaling} to reduce the number of model parameters.\\

Although temporal Markov models have been used to accurately model transport in porous media, there
is still a valid argument against the applicability of these models in the presence of very low
velocities and the verification of the Markov property for discrete temporal Markov models. Here we
address these limitations and study their influence in modeling transport in random networks. 

\section{Particle tracking problem setup}
\label{ptrack}
\begin{figure}[!h]
\begin{center}
\includegraphics[width=0.4\textwidth]{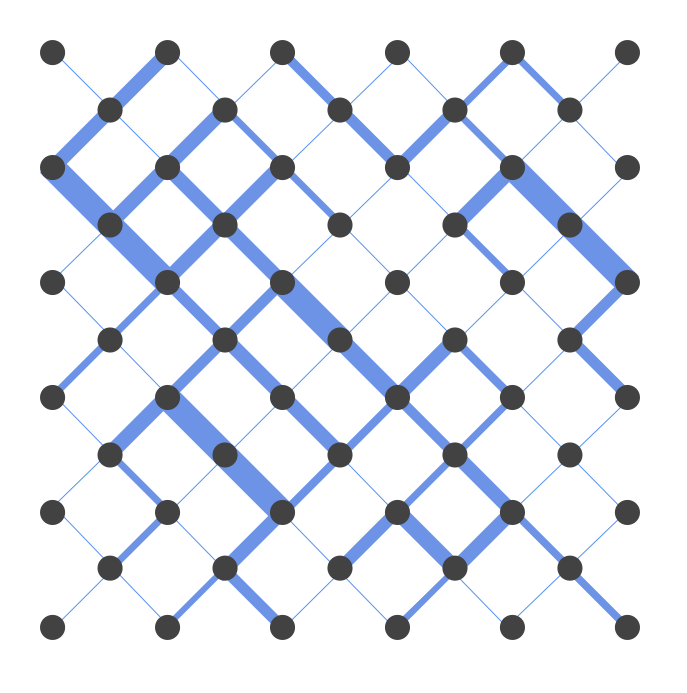}
\caption{Schematic of a porous medium. Adapted from \citet{kang2011spatial}.  \label{figZigzag}}
\end{center} \end{figure}

In this section, a particle tracking setup is described to illustrate the concepts discussed so far.
We study the evolution of particle plumes in structured random networks of the form shown in
Fig.~\ref{figZigzag}. In order to compare the performance of the temporal Markov model with spatial
Markov models, the problem setup is chosen identical to the study performed by
\citet{kang2011spatial}. 
The considered network has $500\times500$ nodes and the link length, $l$, is equal for all links.
All links are oriented at $\pm \pi/4$ radians with respect to the horizontal direction. In each
realization of this structured network
$1000$ tracer particles are injected in the center of the left boundary and tracked in the domain
and $1000$ such realization are considered. 
No-flow boundary conditions are set on the top and bottom of the domain, and a mean pressure
gradient is
imposed by setting the pressure at the left and right boundaries. Each realization of the medium is
obtained by drawing the transmissibility of each link independently from a log-normal distribution
with variance $\sigma^2 = 5$.
The flow problem is first solved for each realization and the velocity field is obtained. We then
simulate the transport of a passive tracer using particle tracking. Here, diffusion inside the links
is neglected and perfect
mixing is assumed inside the nodes. At each node, the particle randomly chooses a link carrying flux
out of that node with a probability proportional to the flux in that link. \\

The outputs of this particle tracking procedure are $x_n^{(i)}$ and $y_n^{(i)}$ and
$t_n^{(i)}$, where $x_n^{(i)}$ and $y_n^{(i)}$ are the position coordinates of particle $i$ after
$n$
transitions and $t_n^{(i)}$ is the elapsed time to get to that position. Given these trajectories
for all particles, we can obtain the ensemble concentration of the contaminant at a given time less
than $\min_i \max_n t_n^{(i)}$, which is the largest time where none of the particles has left the
domain. Alternatively, one can obtain the distribution of the first passage time (FPT) for a certain
$x_t$ plane.\\

To describe the average movement of the particle plume we analyze the Lagrangian statistics of the
particles along each trajectory. As described in the previous section, we assume the velocity
process is stationary and model the velocity of the particles, $\mathbf{v}_n^{(i)}$,  by a Markov
process.  One can use the velocities obtained directly from the MC transport simulations
$(x_n^{(i)}, 
y_n^{(i)}, t_n^{(i)})$. That is,

\begin{equation}
\begin{gathered}
\mathbf{v}_n^{(i)} = [v_x^{(i)}, v_y^{(i)}]_n^{T},\\
v_x^{(i)} = \frac{x_{n+1}^{(i)} - x_{n}^{(i)}}{t_{n+1}^{(i)} - t_{n}^{(i)}}
\text{ and  } v_y^{(i)} = \frac{y_{n+1}^{(i)} - y_{n}^{(i)}}{t_{n+1}^{(i)} - t_{n}^{(i)}}.
\end{gathered}
\label{eqVSpace}
\end{equation}
A Markov model based on these velocities would allow us to efficiently march particles with length
steps equal to $l$. This would
correspond to a spatial Markov model. One can imagine that $l$ is typically much
smaller than the length scales of practical interest, and we might not be interested in the details
of the particle path after each transition to a new node in the network. In the next section the
proposed 
temporal Markov model is discussed.\\

\section{Stencil method: the temporal Markov model}
A temporal Markov model for the velocity process described in the previous section would require
computing the average particle velocity in a sequence of given time intervals. If the averaging time
step is
larger than the mean time required for one transition, by combining multiple transitions together,
averaging will increase the numerical efficiency of the temporal Markov model. Moreover, there is no
distinction between structured and unstructured networks when obtaining the average velocity
process; therefore, averaging would also generalize the model to unstructured networks. We refer to
the averaging time step, $\Delta t_s$, as the stencil time. \\

We represent the average velocity $\overline{\mathbf{v}}_n^{(i)}$ over a time period of $\Delta t_s$
in polar coordinates by its magnitude and the angle $\overline{\theta}_n^{(i)}$ between its
direction and the unit vector in the  $x$ direction. Here the subscript $n$ refers to the $n$'th
time step. The average velocity and average angle are divided into discrete classes as follows:

\begin{equation}
\begin{gathered}
v \in \cup_{j=1}^{n_v} (log(\overline{v})_j, log(\overline{v})_{j+1})\\ 
\text{and } \theta \in \cup_{j=1}^{n_{\theta}} (\overline{\theta}_j, \overline{\theta}_{j+1}),
\end{gathered}
\label{classDef}
\end{equation}
where $n_v$ and $n_{\theta}$ are the number of velocity magnitude and angle classes. 
As previously suggested in \citep{meyer2010particle, kang2011spatial}, in order to better represent
slow transitions we use the logarithm of velocity magnitude for the definition of classes.  The
state of a particle velocity is defined by a $({v}, {\theta})$-pair. We refer to
this temporal model as the velocity-angle stencil method, or the stencil method.\\

Similar to the polar Markovian velocity process (PMVP) ~\citep{meyer2013fast, dunser2016predicting},
the velocity-angle stencil is based on polar coordinates. Unlike the PMVP, the velocity and angle
processes are not considered separately, i.e., they are jointly modeled by one stochastic process in
time. Moreover, the PMVP is a combination of a spatial and a temporal SDE. The stencil method is a
discrete temporal Markov chain.  The advantage of using a
discrete Markov chains is that no assumptions are needed for the functional form of the transition
probabilities. \\

\begin{figure}[!h]
\begin{center}
\includegraphics[width=9.17cm]{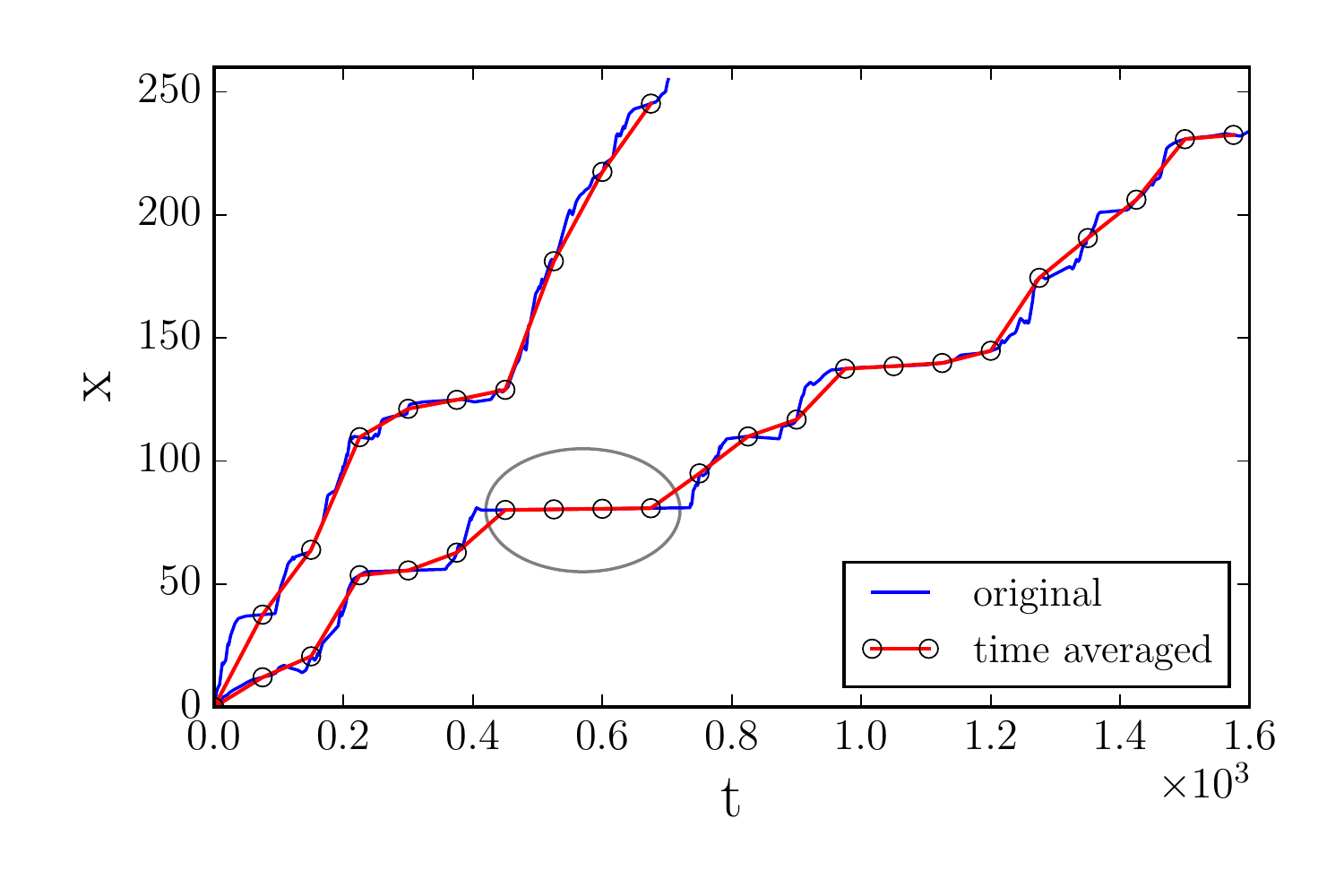}
\caption{Sample trajectory of two particles. Solid blue: original particle path, markers:
trajectory resulting from average velocities. The circled area illustrates an example of a particle
experiencing a relatively low velocity.\label{figTraj}}
\end{center} \end{figure}

Compared to the correlated spatial models that follow every particle transition on the network
(e.g., the one by \citet{kang2011spatial}), the velocity-angle stencil
method has two limitations. First, since we are using average trajectories, collisions (velocity
transitions) do not coincide with the nodes of the underlying physical network. In order to inform
the model about the exact location of the collision, a new variable, $\tau$, can be added to track
the time between the end of every averaging instant and the next collision event.  This idea has
been
discussed in \citet{jenny2016non} and it is not discussed further here.  Another limitation of the
velocity-angle stencil is its inaccuracy in case of very slow transitions, i.e., if $\Delta
t_n^{(i)} = t_{n+1}^{(i)} - t_{n}^{(i)} \gg \Delta t_s$.  This has been noted previously in
\citep{le2008lagrangian, meyer2016testing}. Averaging such time steps would result in many
consecutive steps where the velocity state would remain unchanged. However, using the velocity-angle
stencil, the particle is allowed to change its velocity after every $\Delta t_s$, and this could
limit the persistence of the low velocities and make the model less accurate. Figure~\ref{figTraj}
shows the trajectory of two sample particles and the averaged trajectories for the same particles
for the same number of jumps.  The circled section illustrates an example of a particle experiencing
a relatively low velocity.\\ 

Next, we discuss a method for overcoming the limitation due to very slow transition by enriching the
state space of the stencil model. When averaging the particle trajectories with a stencil time step
$\Delta t_s$, the number of repetitions, denoted by $f$, during a time step is known. This
information can be added to the definition of the particle state. By having the collision frequency
in the state definition, it becomes possible to accurately model particles that stay in a
low-velocity
state for long time intervals. With this new state definition, the attributes of the particle state
are $({v},{\theta}, f)$.  We refer to this model as the extended velocity-angle stencil method or
the extended stencil method.

\section{Model calibration and validation of the Markov property}
Similar to \citet{kang2011spatial}, we use equal probability bins for the log-velocity classes. Due
to averaging, the angle process takes continuous values in $[-\pi, \pi]$ and will include values
that do not coincide with the initial network link directions. We use equal-width bins for the
velocity angle classes.  In the extended stencil method, for each observed frequency in the input
data, a separate class was allocated for the corresponding $(v, {\theta})$ pair. We refer to the
discrete transition matrix for the stencil method as $p_m(v, {\theta}|v', {\theta}')$. This is the
probability of encountering the state $(v,{\theta})$ after $m$ transitions, assuming that we started
from the state $(v',{\theta}')$. Similarly $p_m(v, {\theta}, f|v', {\theta}', f')$ is the discrete
transition matrix for the extended stencil method.  Both transition matrices were obtained by
counting the observed transitions in the particle tracking simulation results, which corresponds to
the maximum likelihood estimation of the transition probabilities. We also define the aggregate
velocity transition matrix $T_m^{v} (i,j)$, which defines the probability of encountering velocity
class $i$ after $m$ transitions starting from velocity class $j$, and the aggregate angle transition
matrix $T_m^{\theta} (i,j)$, which is defined in a similar way. The mean transition time observed in
the particle tracking data, $\overline{\delta t}$, is considered as a time scale.\\

\begin{figure*}[!h]
\begin{center}
\includegraphics[width=\textwidth]{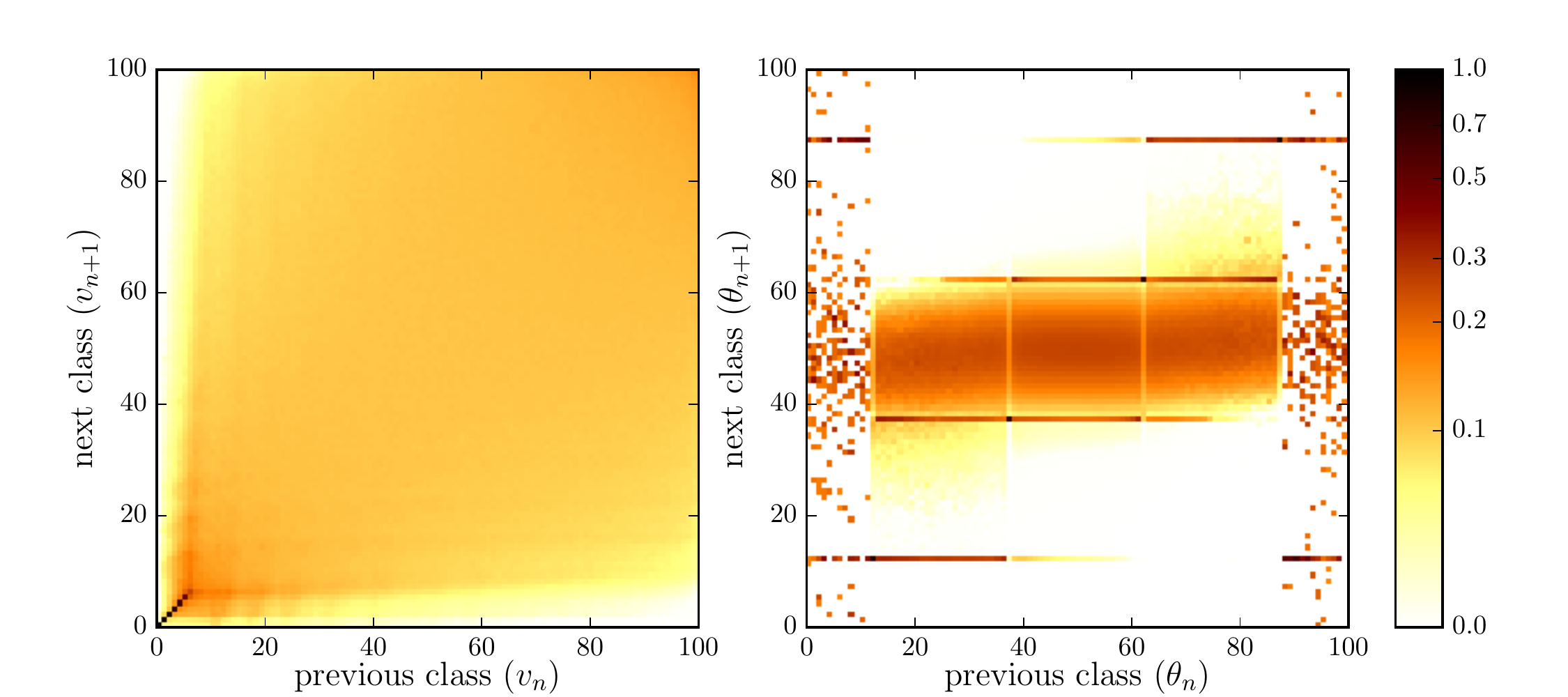}
\caption{Left: aggregate velocity transition matrix for $m=1$, $T_1^v (i,j)$; right: aggregate angle
transition matrix for $m=1$, $T_1^{\theta} (i,j)$ for the velocity-angle stencil for $\Delta t_s =
10\overline{\delta t}$. The square root of the values are plotted in logarithmic scale. 
\label{transMatStencil}}
\end{center} \end{figure*}

\begin{figure*}[!h]
\begin{center}
\includegraphics[width=\textwidth]{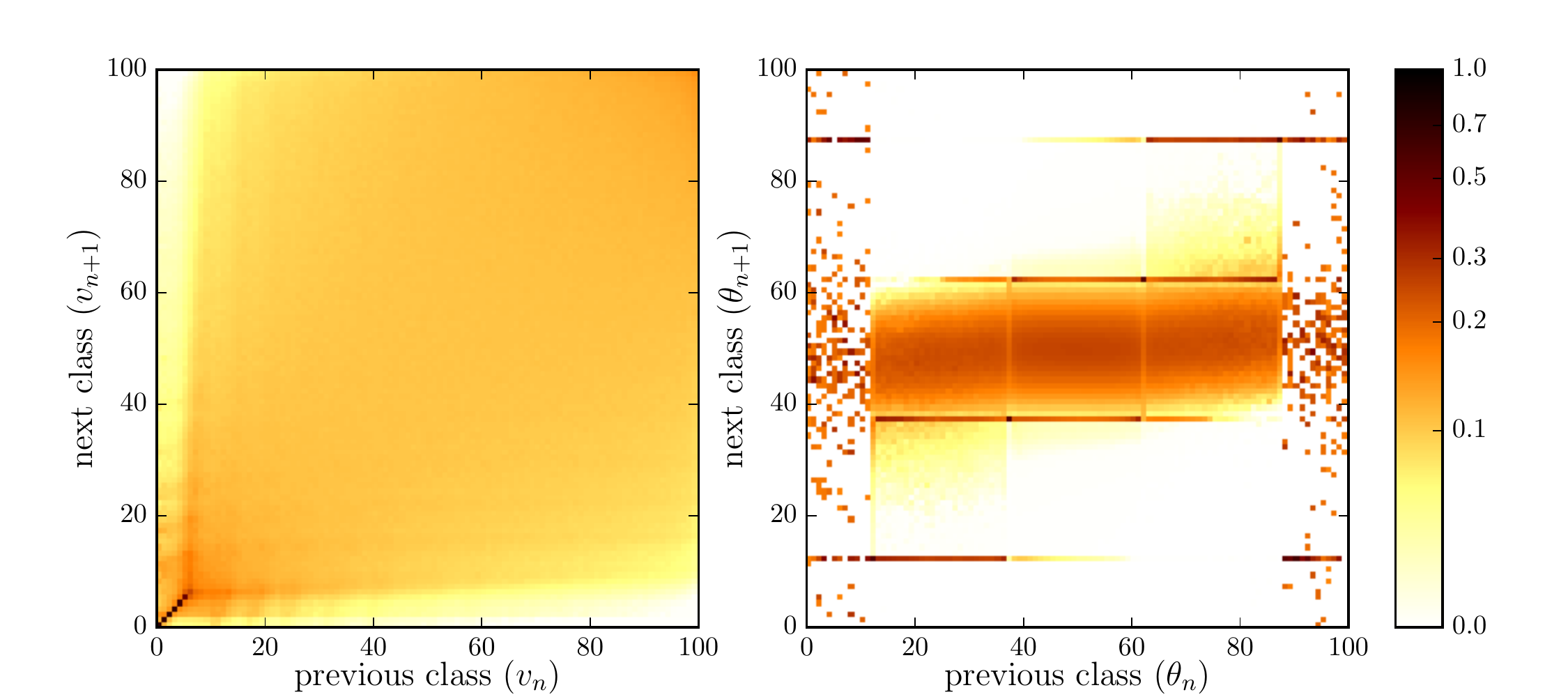}
\caption{Left: aggregate velocity transition matrix for $m=1$, $T_1^v (i,j)$; right: aggregate angle
transition matrix for $m=1$, $T_1^{\theta} (i,j)$ for the extended velocity-angle stencil for
$\Delta t_s = 10\overline{\delta t}$. The square root of the values are plotted in logarithmic
scale. 
\label{transMatExtended}}
\end{center} \end{figure*}

\begin{figure*}[!h]
\begin{center}
\includegraphics[width=\textwidth]{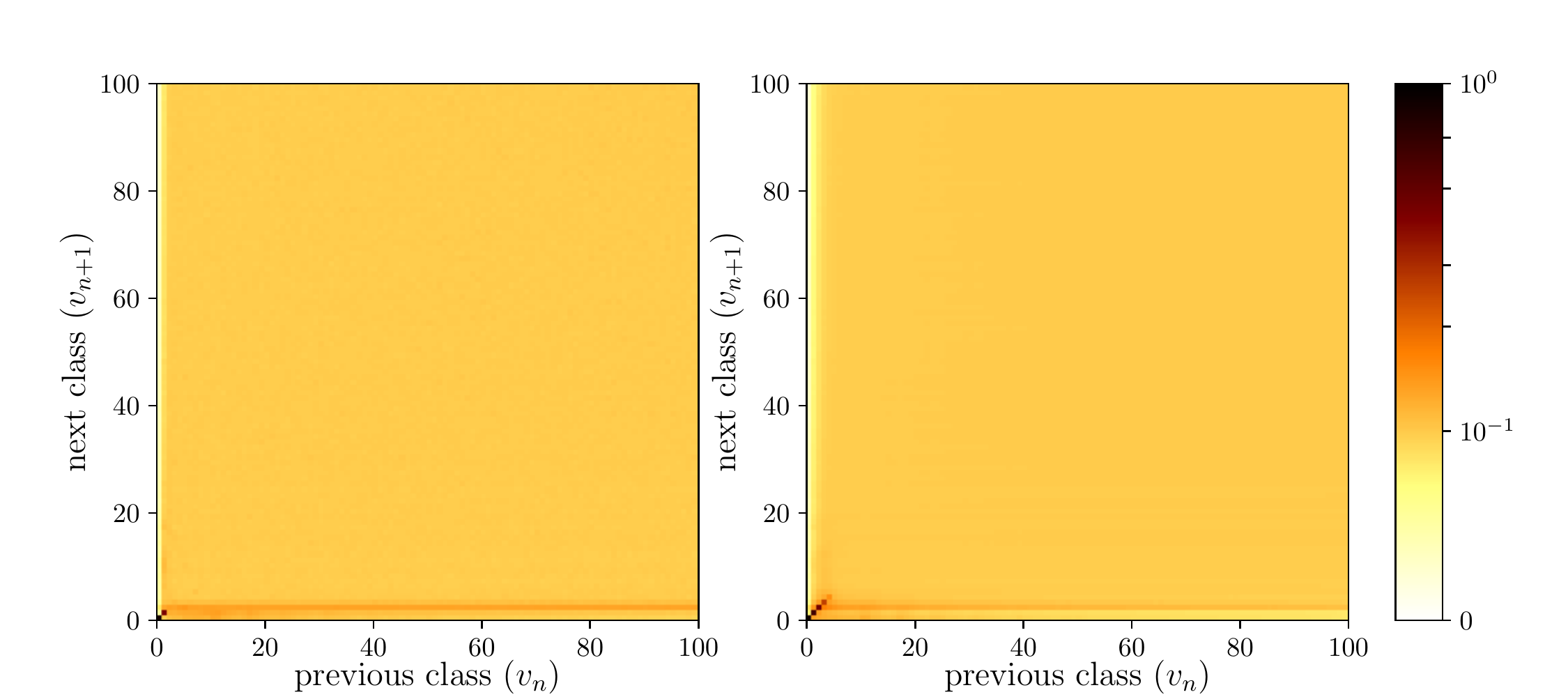}
\caption{Verification of the Chapman-Kolmogorov relation for velocity-magnitude classes for the
stencil method for $\Delta t_s = 10\overline{\delta t}$; left: $T_5^v(i,j)$; right:
${T_1^v(i,j)}^5$. The square root of the values are plotted in logarithmic scale. 
\label{matVstencilCompare}}
\end{center} \end{figure*}

\begin{figure*}[!h]
\begin{center}
\includegraphics[width=\textwidth]{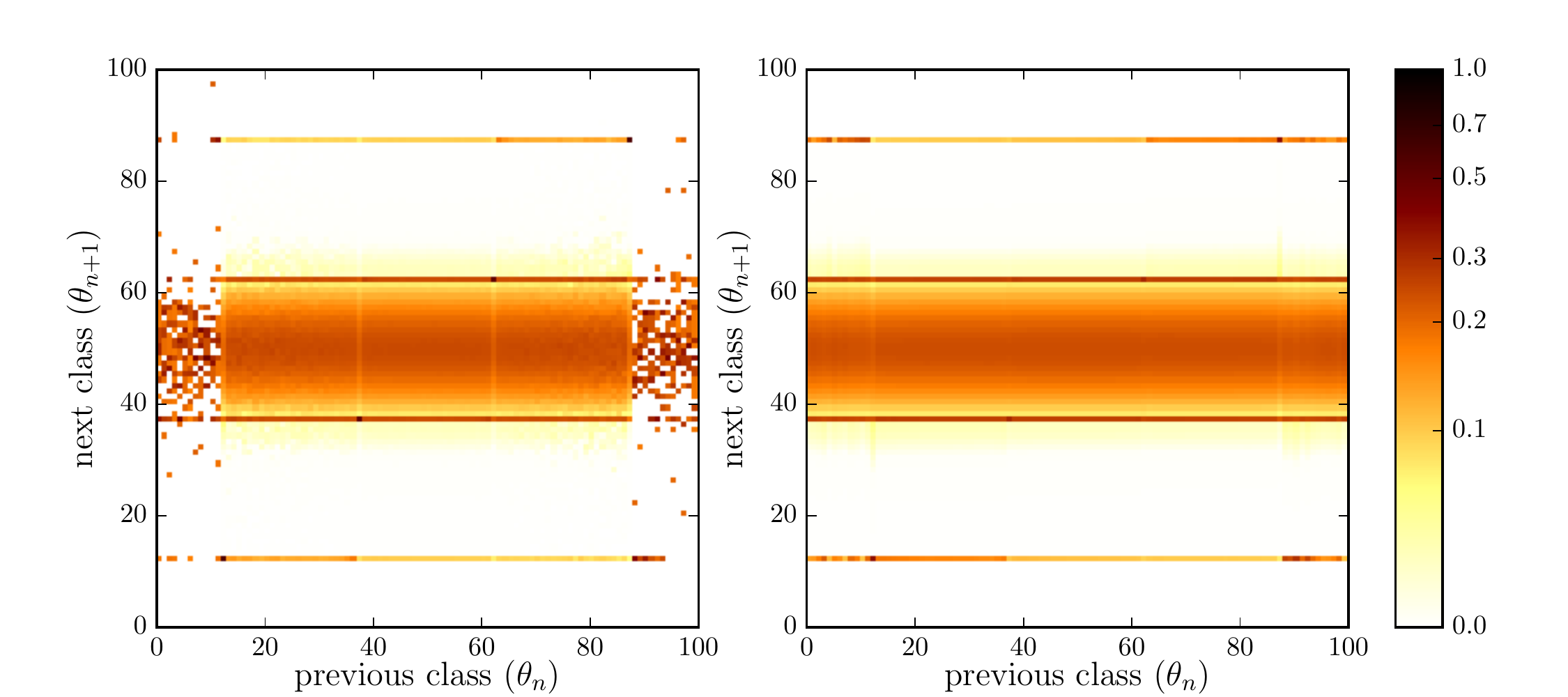}
\caption{Verification of the Chapman-Kolmogorov relation for the velocity-angle classes for the
stencil method for $\Delta t_s = 10\overline{\delta t}$; left: $T_5^{\theta} (i,j)$; right:
${T_1^{\theta}(i,j)}^5$. The square root of the values are plotted in logarithmic scale. 
\label{matTstencilCompare}}
\end{center} \end{figure*}

\begin{figure*}[!h]
\begin{center}
\includegraphics[width=\textwidth]{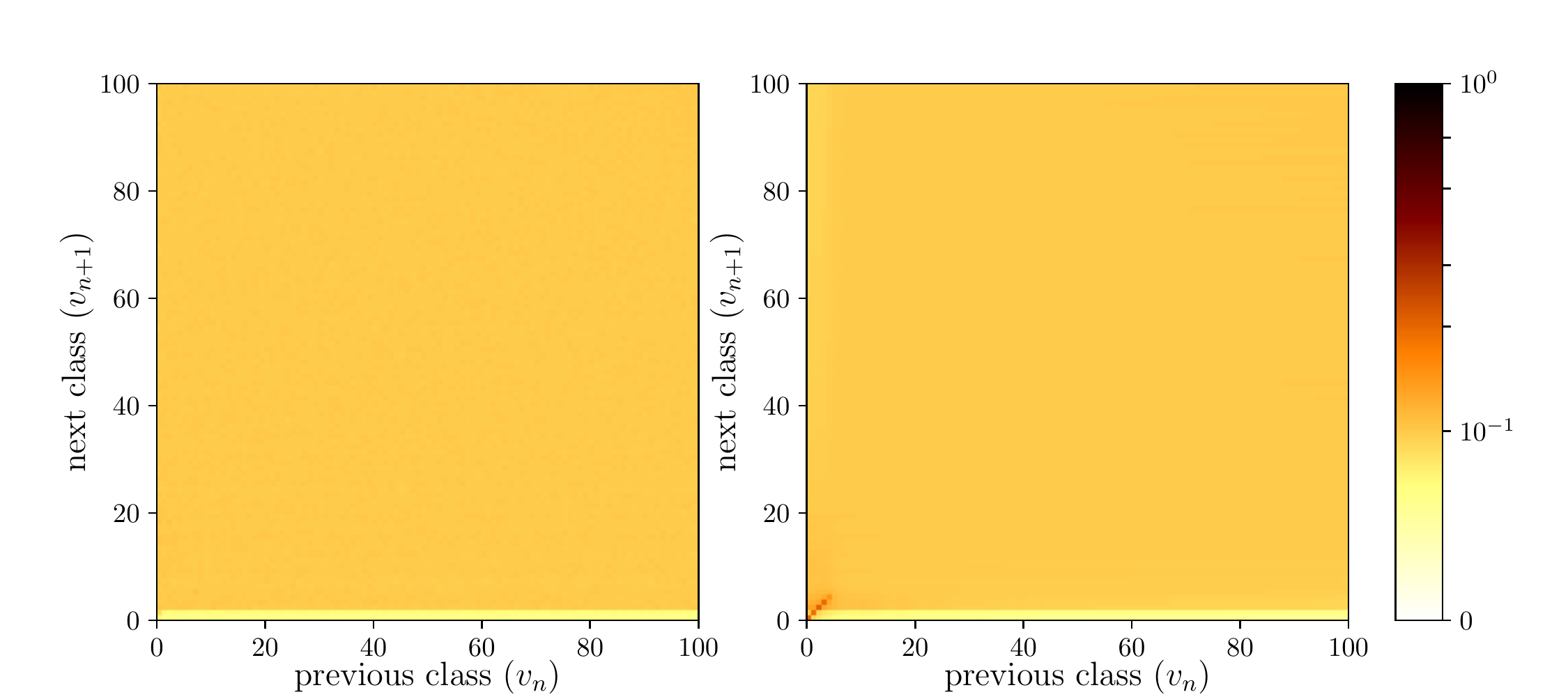}
\caption{Verification of the Chapman-Kolmogorov relation for velocity-magnitude classes for the
extended stencil method for $\Delta t_s=10\overline{\delta t}$; left: $T_5^v(i,j)$; right:
${T_1^v(i,j)}^5$. The square root of the values are plotted in logarithmic scale. 
\label{matVextendedCompare}}
\end{center} \end{figure*}

\begin{figure*}[!h]
\begin{center}
\includegraphics[width=\textwidth]{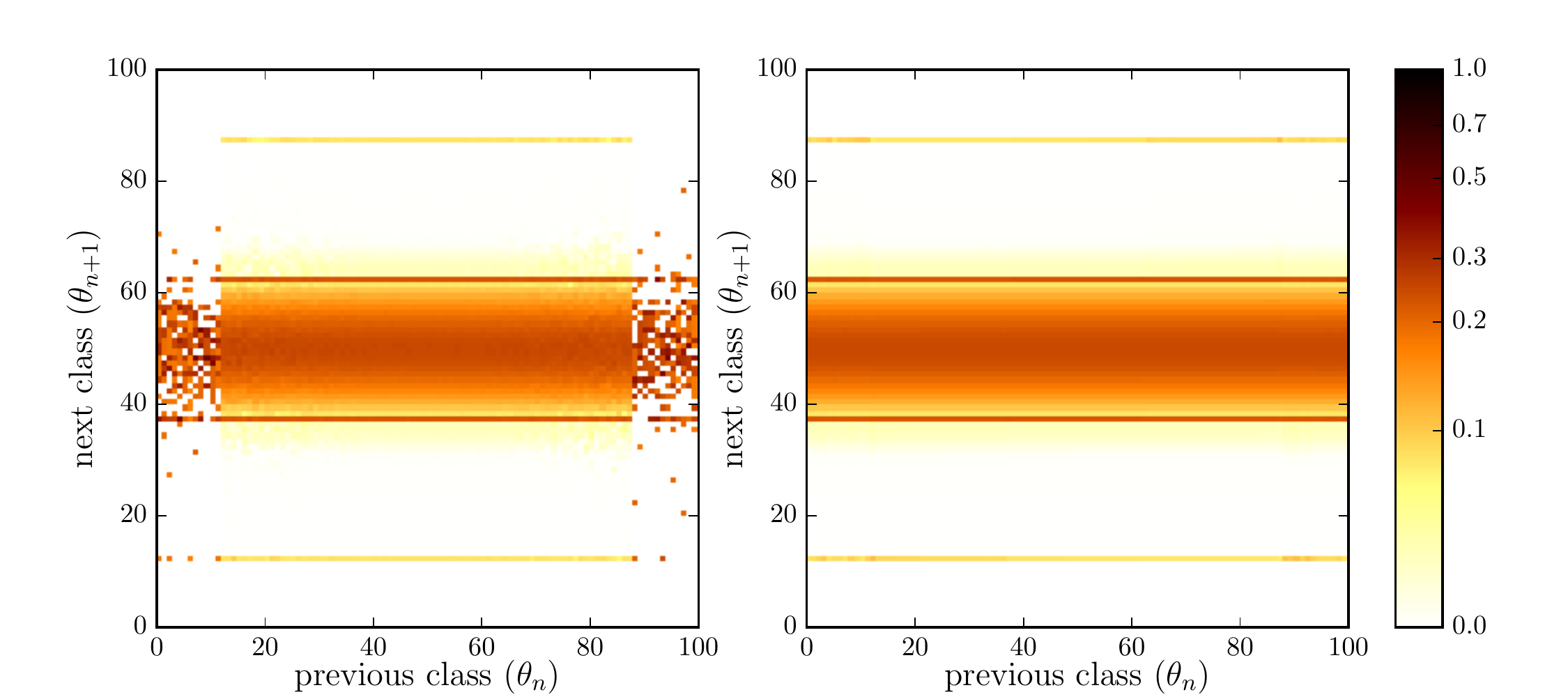}
\caption{Verification of the Chapman-Kolmogorov relation for the velocity-angle classes for the
extended stencil method for $\Delta t_s=10\overline{\delta t}$; left: $T_5^{\theta} (i,j)$; right:
${T_1^{\theta}(i,j)}^5$. The square root of the values are plotted in logarithmic scale. 
\label{matTextendedCompare}}
\end{center} \end{figure*}

The aggregate velocity and angle transition matrices for one transition with $\Delta t_s =
10\overline{\delta t}$ for the stencil method and the extended stencil method are shown in
Figs.~\ref{transMatStencil} and~\ref{transMatExtended}. The transition matrices for these two models
will be different, since repeating $(v, \theta)$ pairs are counted as a transition to the same state
in the stencil method, which is not the case for the extended velocity-angle stencil. The resulting
aggregate transition matrices are almost identical except for the lower left corner of the velocity
transition matrices. Provided that a Markov process can closely model transitions between different
states, we expect that the Chapman-Kolmogorov relation \citep{resnick2013adventures} holds for
these transitions.  We perform a test to compare the m-step aggregate transition matrices for $v$
and $\theta$ with the m-fold product of the corresponding one-step transition matrices.  The
aggregate five-step velocity transition matrix $T_5^v (i,j)$ and the aggregate five-step angle
transition matrix $T_5^{\theta} (i,j)$ are compared to ${T_1^v(i,j)}^5$ and ${T_1^{\theta}(i,j)}^5$
from the stencil method in Figs.~\ref{matVstencilCompare} and \ref{matTstencilCompare} for $\Delta
t_s = 10\overline{\delta t}$. The same comparison is performed for the extended-stencil method and
presented in Figs.~\ref{matVextendedCompare} and \ref{matTextendedCompare}. As can be seen from
these four figures, the Chapman-Kolmogorov relation approximately holds for all angles and for
velocity classes with $j>5$.\\

\begin{figure*}[!h]
\begin{center}
\includegraphics[width=\textwidth]{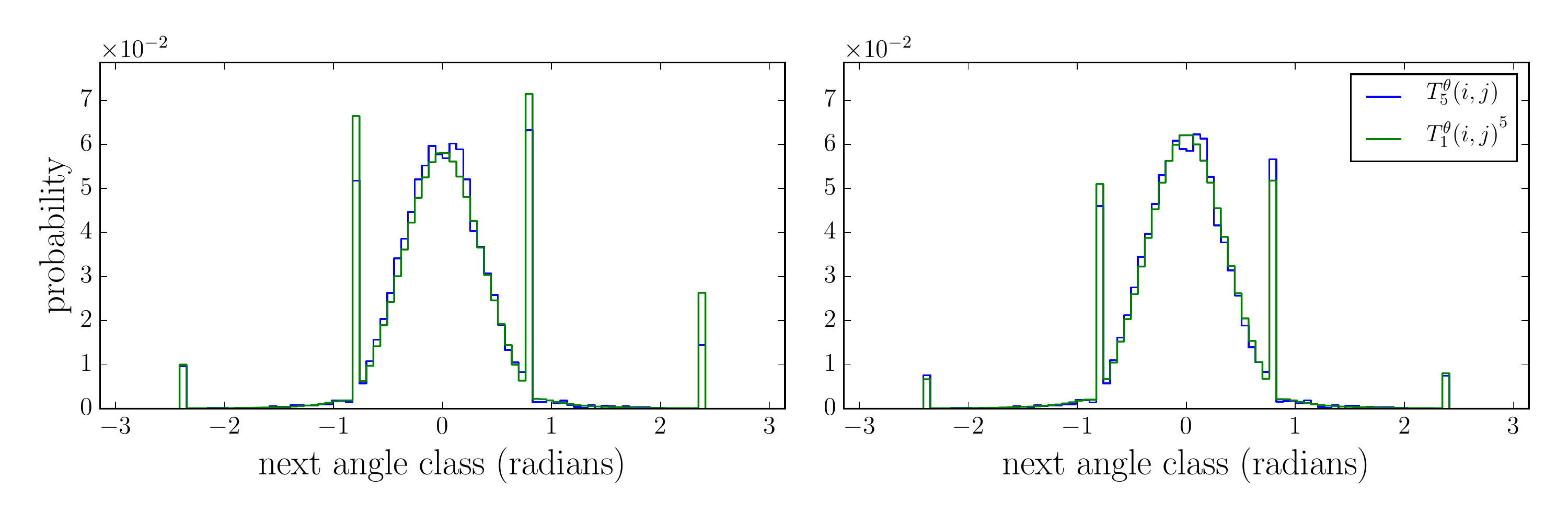}
\caption{Column-wise comparison of $T_5^{\theta} (i,j)$ and ${T_1^{\theta}(i,j)}^5$ for an initial
angle in the second quarter with $j=78$ for $\Delta t_s = 10\overline{\delta t}$; left: stencil
method; right: extended stencil method.
\label{t_j78_both}}
\end{center} \end{figure*}

\begin{figure*}[!h]
\begin{center}
\includegraphics[width=\textwidth]{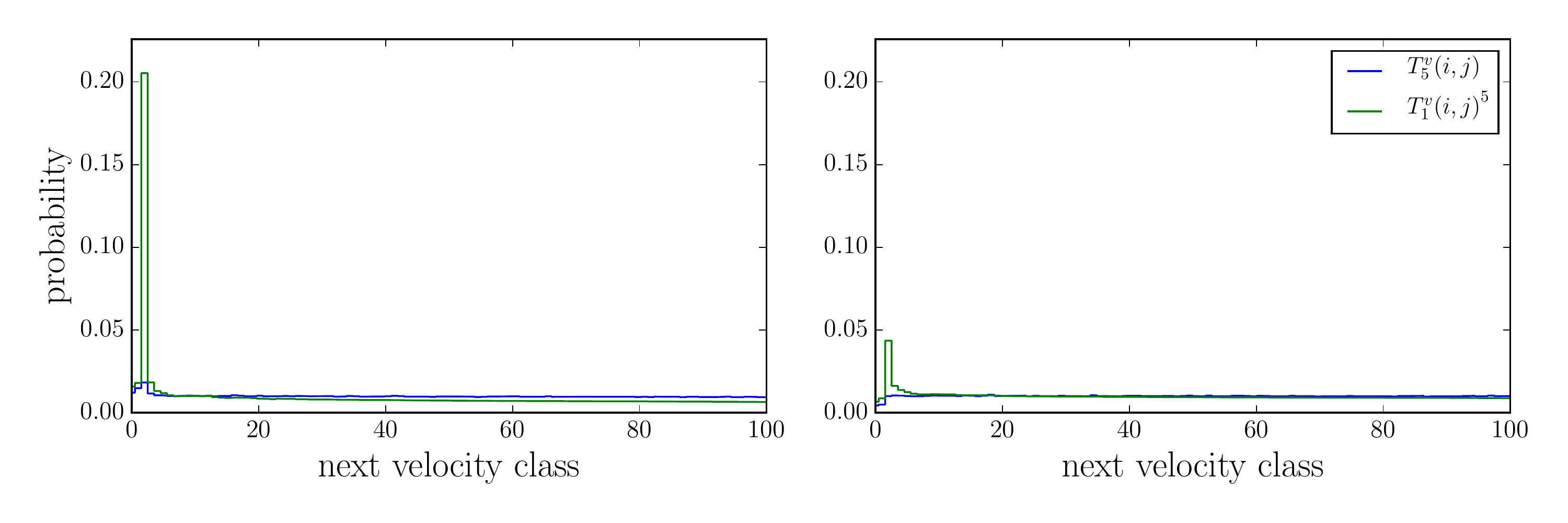}
\caption{Column-wise comparison of $T_5^v(i,j)$ and ${T_1^v(i,j)}^5$ for a slow initial velocity
class with $j=2$ for $\Delta t_s = 10\overline{\delta t}$; left: stencil method; right: extended
stencil method.
\label{v_j2_both}}
\end{center} \end{figure*}

\begin{figure*}
\begin{center}
\includegraphics[width=\textwidth]{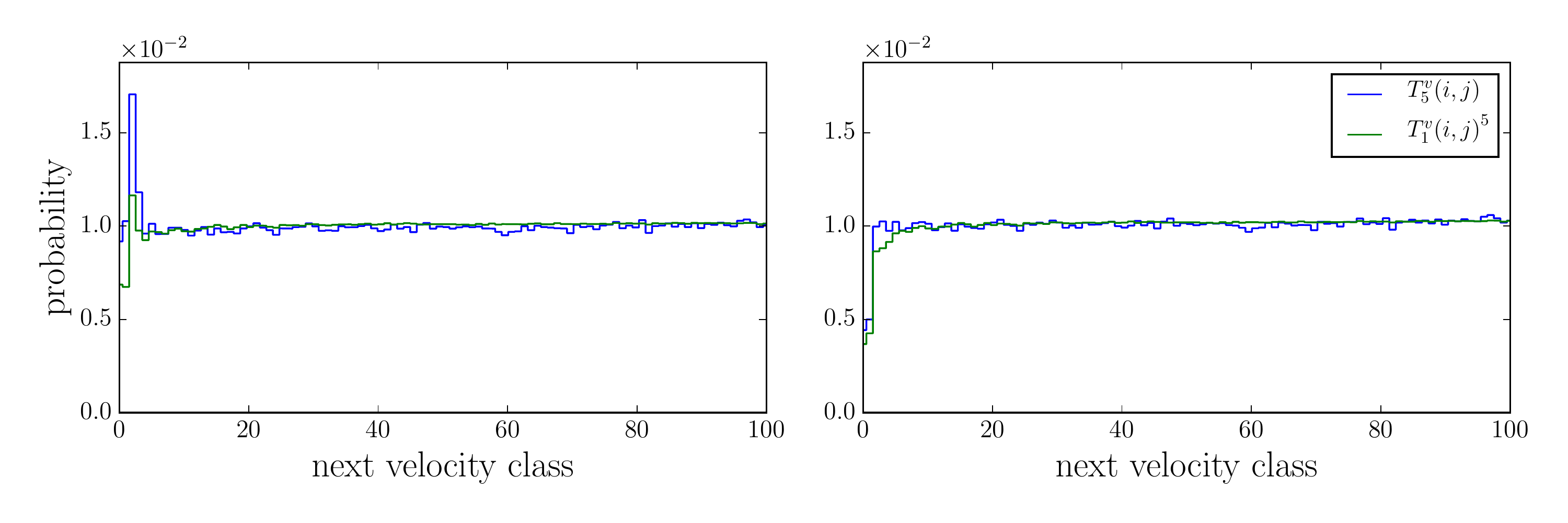}
\caption{Column-wise comparison of $T_5^v(i,j)$ and ${T_1^v(i,j)}^5$ for a fast initial velocity
class with $j=98$ for $\Delta t_s = 10\overline{\delta t}$; left: stencil method; right: extended
stencil method.
\label{v_j98_both}}
\end{center} \end{figure*}

A column-wise comparison of the aggregate angle transition matrices for an angle in the second
quarter ($j=78$) is shown in Fig.~\ref{t_j78_both}. Since the mean-flow directions is from left to
right ($\theta=0$), there are not that many sample paths with velocity angles close to $\pi$ or
$-\pi$. This explains the noise observed in the aggregate angle transition matrices of
Figs.~\ref{transMatStencil} and~\ref{transMatExtended}. The column-wise comparison of the aggregate
velocity transition matrices for a low velocity class ($j=2$) and a high velocity class ($j=98$) are
shown in Figs.~\ref{v_j2_both} and~\ref{v_j98_both}. Although there is a clear deviation from
Markovianity for both the stencil method and the extended stencil method, these results suggest that
enriching the state space of the Markov model would result in significantly better predictions of
transitions from extremely low velocity values. This is best illustrated in Fig.~\ref{v_j2_both}.
The extended velocity-angle stencil also improves the predicted lag five transitions from a fast
velocity class to slow velocity classes as seen in Fig.~\ref{v_j98_both}.\\


Similar comparisons were performed for larger stencil times, $\Delta t_s > 10\overline{\delta t}$,
for both
the stencil method and the extended stencil method. For these larger stencil times both models lead
to
smaller errors in predicting the five-step transition probabilities. This is expected, since with
larger stencil times we are combining more transitions and consecutive transitions 
become less correlated and hence less challenging to predict for the Markov models. In the limit of
$\Delta t_s \rightarrow \infty$ we expect the average velocity distribution to converge to an
equilibrium distribution and every column of the transition matrices would be
equal to the corresponding equilibrium distribution. Hence, for very large $\Delta t_s$ the
one-point distribution of $\overline{v}$ would be sufficient for modeling dispersion. The choice of
$\Delta t_s$ would also depend on the temporal resolution of interest. The range of stencil times
where a correlated stencil is required for accurate predictions of contaminant transport is further
discussed in section~\ref{section:results}.\\ 

For $\Delta t_s < 10\overline{\delta t}$, both the velocity-angle stencil and the extended
velocity-angle
stencil lead to errors in predicting the transition probabilities for a larger set of velocity
classes. However, from a computational point of view we are not interested in these stencil times,
since by combining very few transitions, only small speed-ups can be expected compared to
correlated spatial models. In case we are interested in these smaller time steps, the right tool has
already been developed (correlated CTRW). The resulting plumes and first passage time curves
predicted by these smaller stencil times are included in section~\ref{section:results}.\\

The results presented in this section indicate that the transition probabilities of the MC data 
are not strictly Markovian
for any $\Delta t_s$, which is consistent with the findings of \citet{le2008lagrangian}. However, we
do not expect an exact Markovian structure in the MC results to begin with. For example, the MC
results indicate that in every realization there are fast paths between the two ends of the domain,
and knowing the full history of the velocity of a particle would improve our prediction of whether
that particle is in a preferential path or not. This clearly indicates that the velocity process is
not strictly Markovian. Therefore, although verifying the Chapman-Kolmogorov relation gives us
useful intuition about the value of a Markov assumption for this problem, the usefulness of this
assumption needs to be judged by the predictive power of the model and the error induced due to this
assumption. In the next section, numerical results from both stencil models are presented and
compared to the particle ensemble from the particle tracking simulations.

\section{Numerical results}
\label{section:results}
Different velocity-angle stencils and extended velocity-angle stencils were obtained for different
averaging times and the average transport in the network was compared with the predictions of the
stencil methods. The performance of the stencil models were compared to an implementation of the
correlated CTRW method. The code for generating the MC data and performing these comparisons is
available at \url{https://github.com/amirdel/dispersion-random-network}
\citep{delgoshaieMarkovData}. First, we present results for $\Delta t_s =
20\overline{\delta t}$. The results for smaller and larger time steps are discussed afterwards.\\

\begin{figure}
\begin{center}
\includegraphics[width=\figsize]{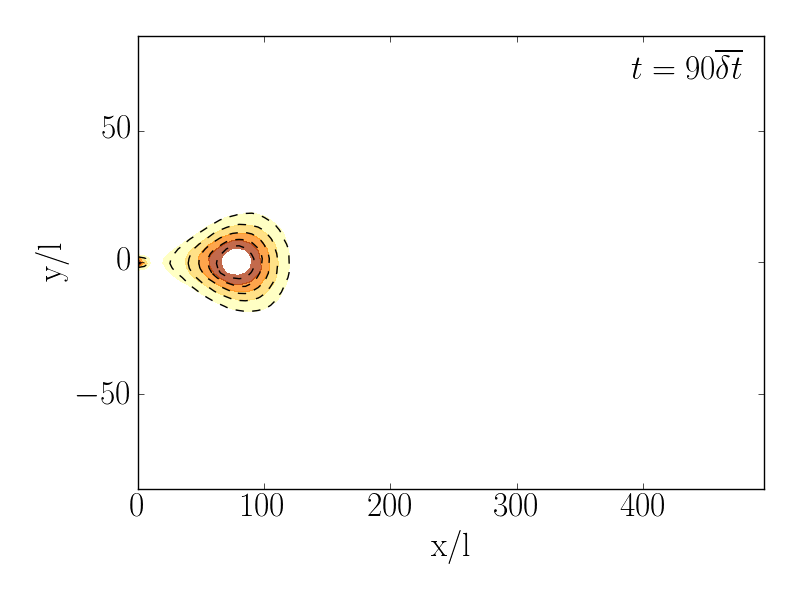}
\caption{Contaminant concentration at non-dimensional time $t / \overline{\delta t} = 90$.
Comparison of the velocity-angle
stencil (dashed lines) with the reference particle tracking data (filled contour map) for the same
contour levels. \label{fig2d_1}}
\end{center} \end{figure}

\begin{figure}
\begin{center}
\includegraphics[width=\figsize]{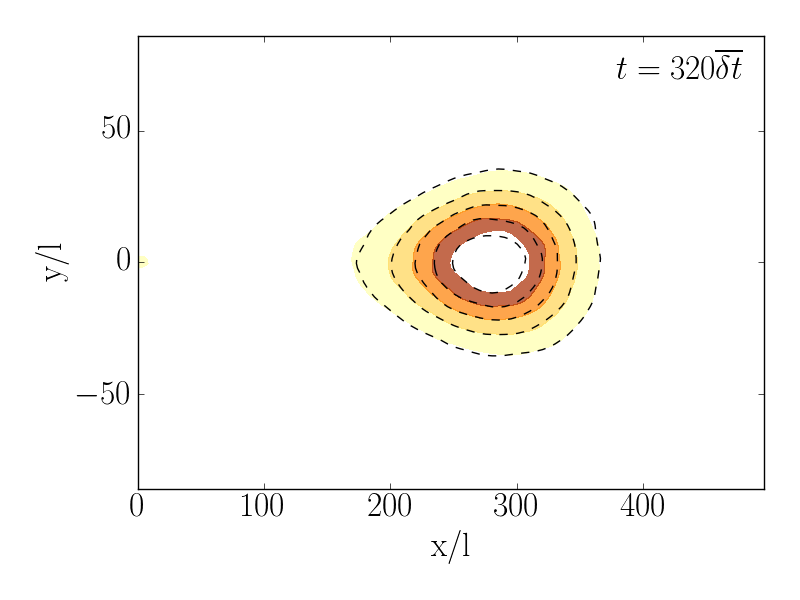}
\caption{Contaminant concentration at non-dimensional time $t / \overline{\delta t} = 320$.
Comparison of the
velocity-angle stencil (dashed lines) with the reference particle tracking data (filled contour map)
for the same
contour levels. \label{fig2d_2}}
\end{center} \end{figure}

\begin{figure*}
\centering
\includegraphics[width=\textwidth]{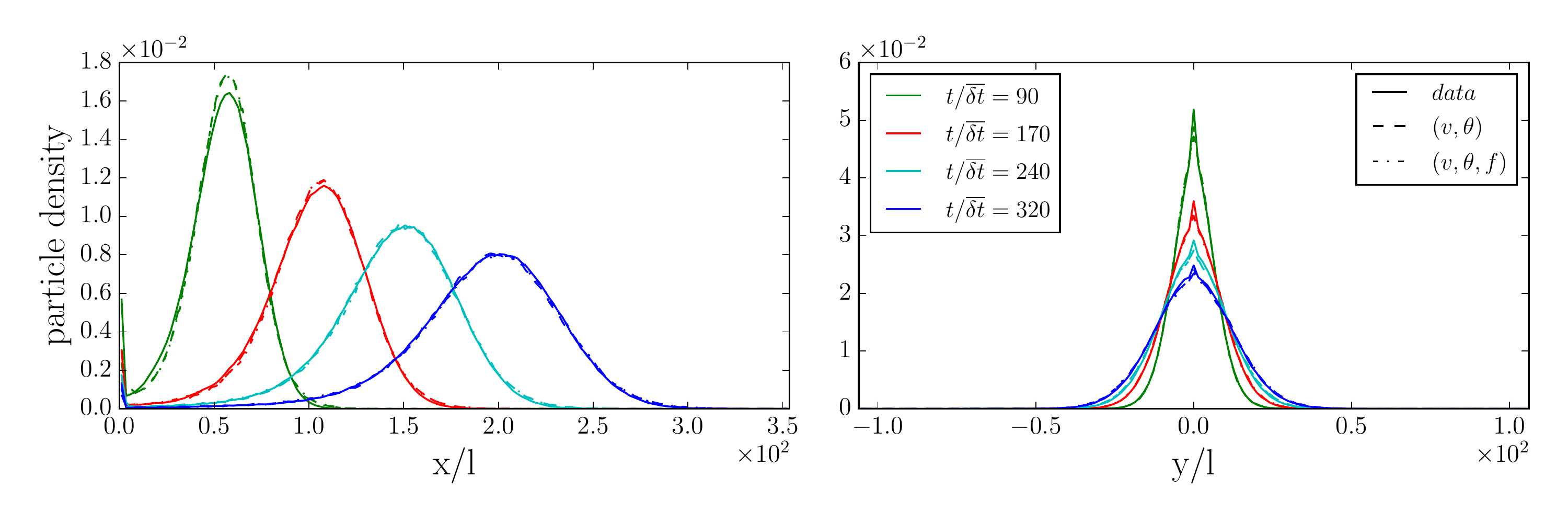}
\caption{Comparison of plume concentration at different times: particle tracking data (solid lines),
velocity-angle stencil (dashed lines), extended velocity-angle stencil (dash dots). Stencil time is 
$20\overline{\delta t}$ for both models.}
\label{3way_compare_s}
\end{figure*}

\begin{figure}
\centering
\includegraphics[width=\figsize]{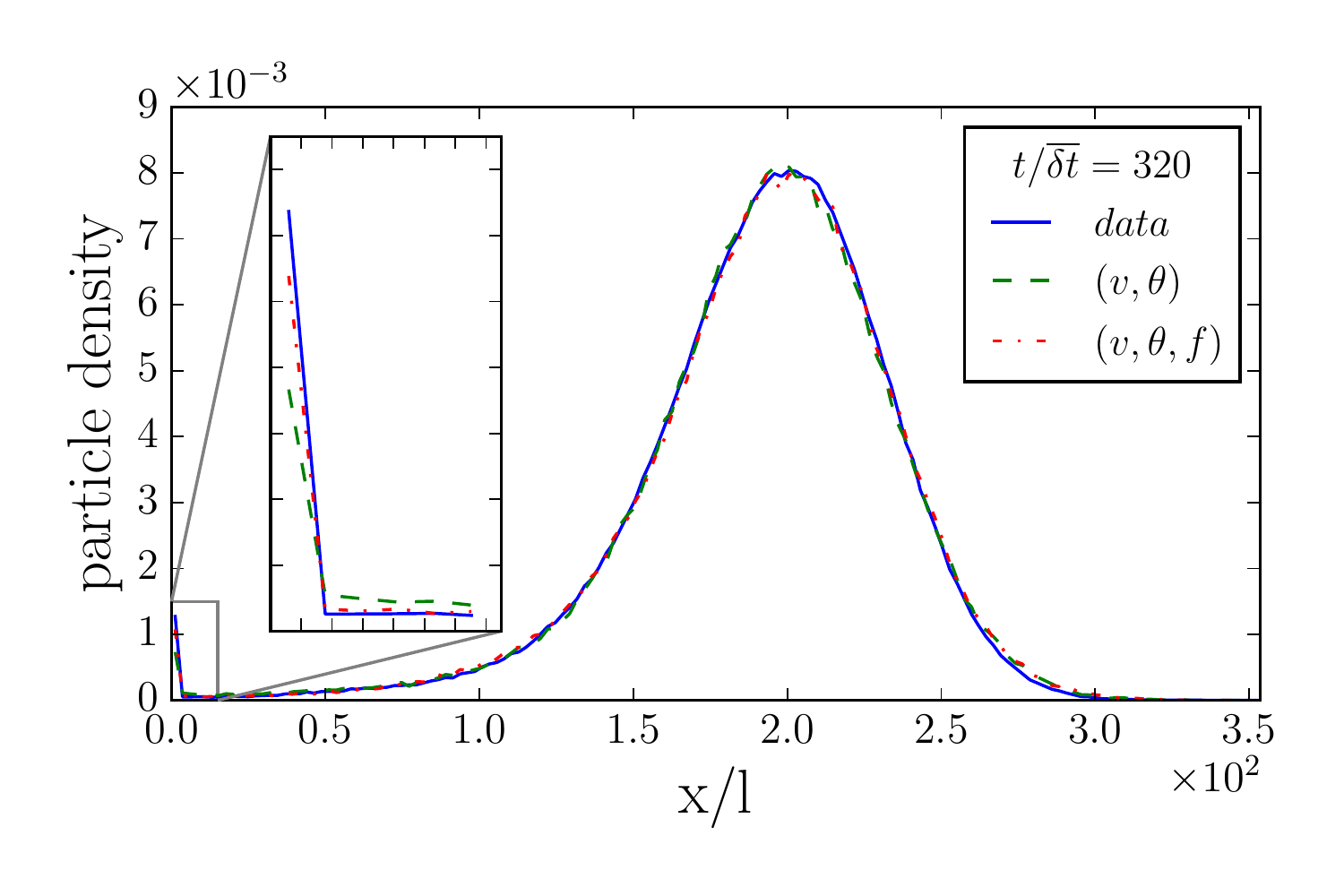}
\caption{Comparison of plume concentration at $t/ \overline{\delta t} = 320$: particle tracking data
(solid lines), velocity-angle stencil (dashed lines), extended velocity-angle stencil (dash dots). 
Inset: zooming on the slow tail of the plume.}
\label{zoom}
\end{figure}

First, we consider the velocity-angle stencil.  Figs.~\ref{fig2d_1} and \ref{fig2d_2} show the
particle plume and the predicted plume by the velocity-angle stencil at dimensionless times $t/
\overline{\delta t} = 90$ and $320$, respectively.  As these
figures suggest, the stencil model captures the distribution of the tracer with great accuracy and
the non-Gaussian characteristics of the plume are well captured using this simpler model.  The same
contaminant plume was also simulated with the extended stencil model. In Fig.~\ref{3way_compare_s}
the results from both models are compared to the reference concentration distributions from the
particle tracking simulations. Both models capture the spreading of the plume accurately for
different times in both longitudinal and transverse directions. When magnifying
the slow tail of the plume in Fig.~\ref{zoom}, it is
observed that the extended stencil captures this slow tail better than the velocity-angle stencil.\\

\begin{figure}
\centering
\includegraphics[width=\figsize]{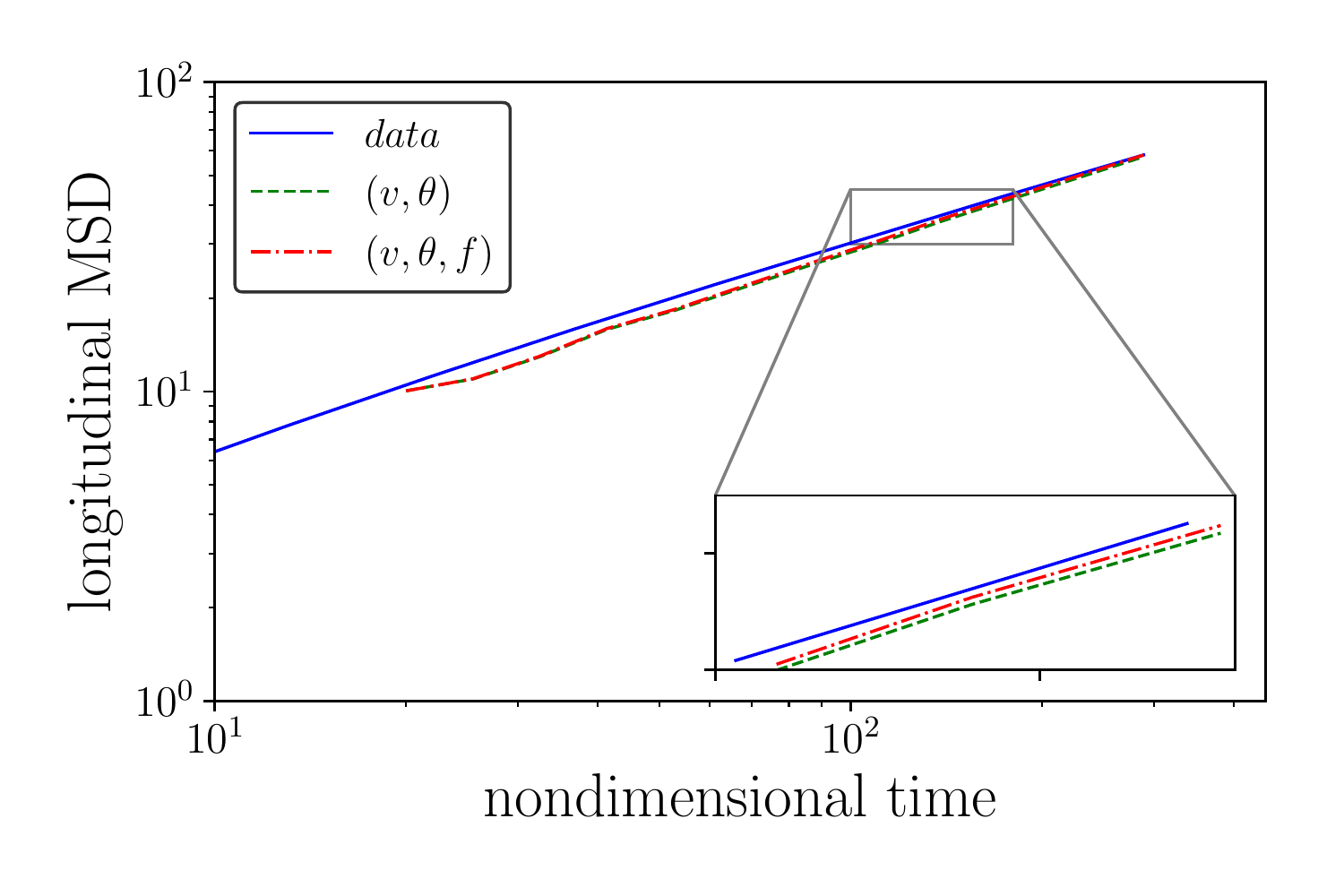}
\caption{Comparison of longitudinal mean square difference of the plume at different times: particle
tracking data (solid lines), velocity-angle stencil (dashed lines), extended velocity-angle stencil 
(dash dots). 
Stencil time is $20\overline{\delta t}$ for both models.}
\label{3way_long}
\end{figure}

\begin{figure}
\centering
\includegraphics[width=\figsize]{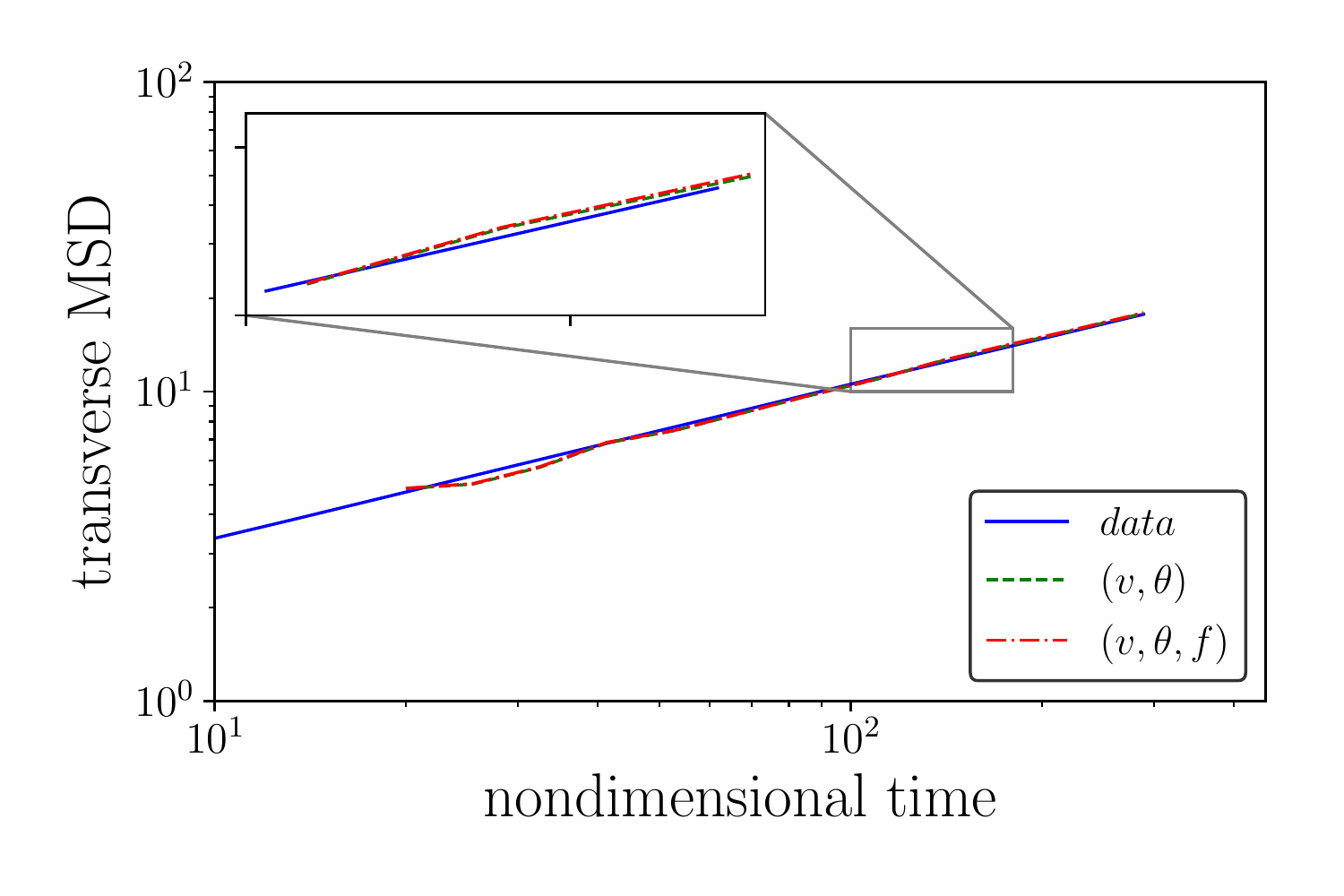}
\caption{Comparison of traverse mean square difference of the plume for different times: particle
tracking data (solid lines), velocity-angle stencil (dashed lines), extended velocity-angle stencil
(dash dots).  Stencil time is $20\overline{\delta t}$ for both models.}
\label{3way_trans}
\end{figure}

The second central moment of the particle plume or the mean square displacement (MSD), with respect
to the plume center of mass in the longitudinal directions obtained from the temporal
Markovian models is compared to the MSD obtained from the ensemble plume in
Fig.~\ref{3way_long}. It can be observed that the predictions from both models are very similar. A
closer inspection (inset of Fig.~\ref{3way_long}) illustrates that the extended stencil method
improves the predictions of the longitudinal MSD. A similar comparison is shown for the transverse
direction
in Fig.~\ref{3way_trans}. The results show that using the extended class definition does not improve
the
predictions of the second moment in the transverse direction.\\

\begin{figure}
\centering
\includegraphics[width=\figsize]{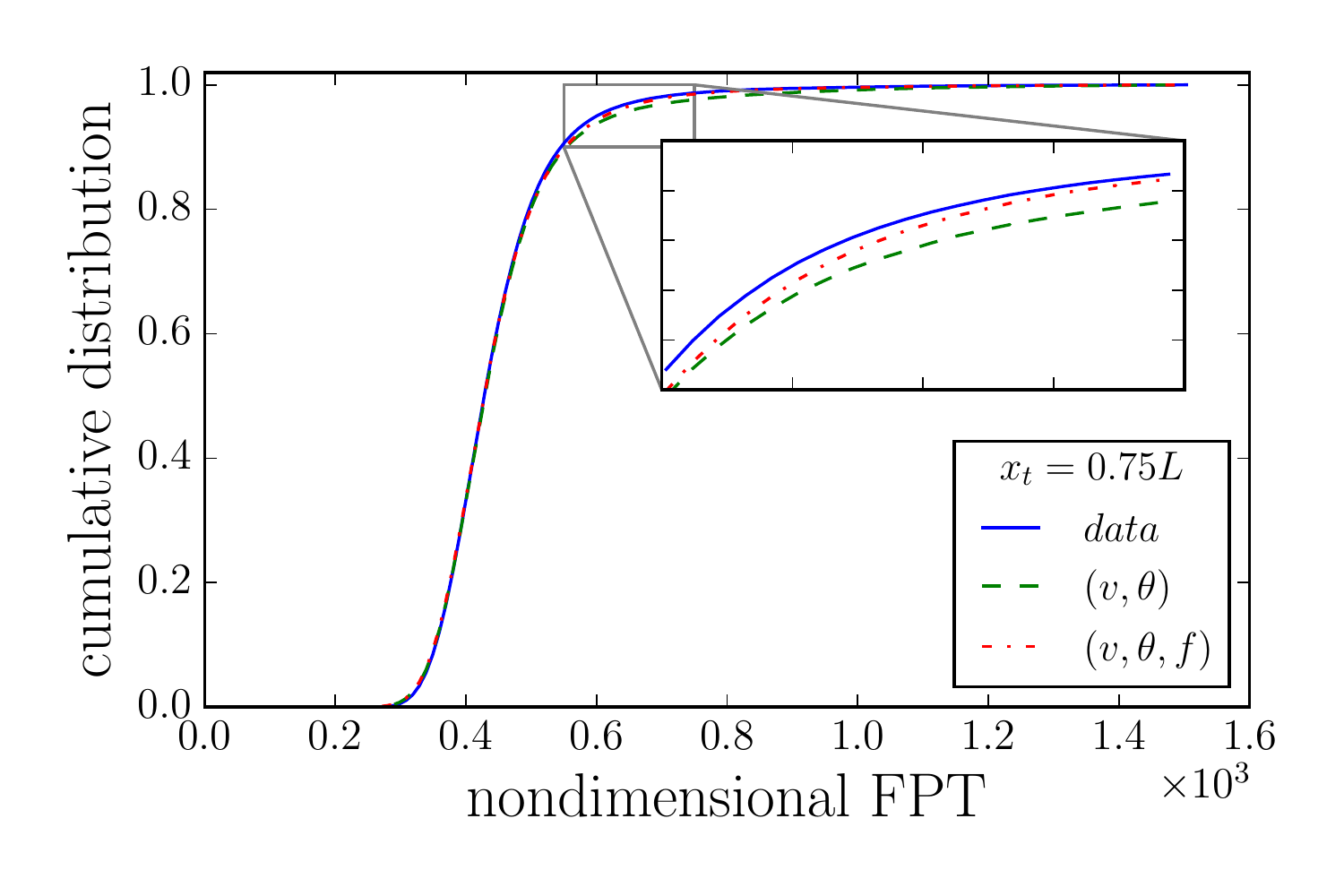}
\caption{First passage time CDF for $x/L = 0.75$: particle tracking data (solid lines),
velocity-angle stencil (dashed lines), extended velocity-angle stencil (dash dots) for $\Delta t_s =
20\overline{\delta t}$.} 
\label{figBt}
\end{figure}

Another important feature of anomalous transport is the long tail in first passage time (FPT)
distributions. Figure~\ref{figBt} shows the comparison of the FPT for non-dimensional length $x_t/L
= 0.75$, where $x_t$ is the target $x$ plane and $L$ is the domain length.  The long tail of the
first passage time CDF is well captured by both stencil models.  This comparison was performed for
$x_t/L = 0.25$, $0.5$ and $0.75$, and the predictions from both models are accurate for all three
planes. A closer inspection of the FPT curve (Fig.~\ref{figBt}) illustrates that the extended class
definition improves the prediction of the FPT distribution.\\

\begin{figure*}
\centering
\includegraphics[width=\textwidth]{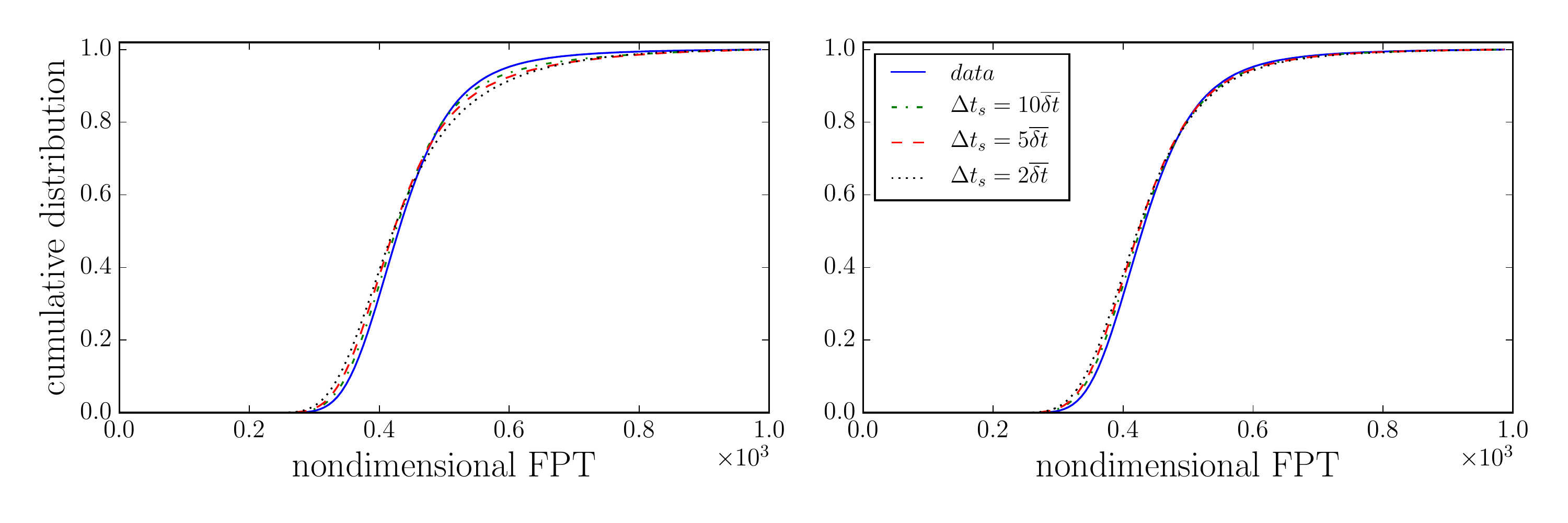}
\caption{First passage time CDF for $x/L = 0.75$ for different stencil times; left: stencil method;
right: extended stencil method.} 
\label{dt10_bt}
\end{figure*}

\begin{figure*}
\centering
\includegraphics[width=\textwidth]{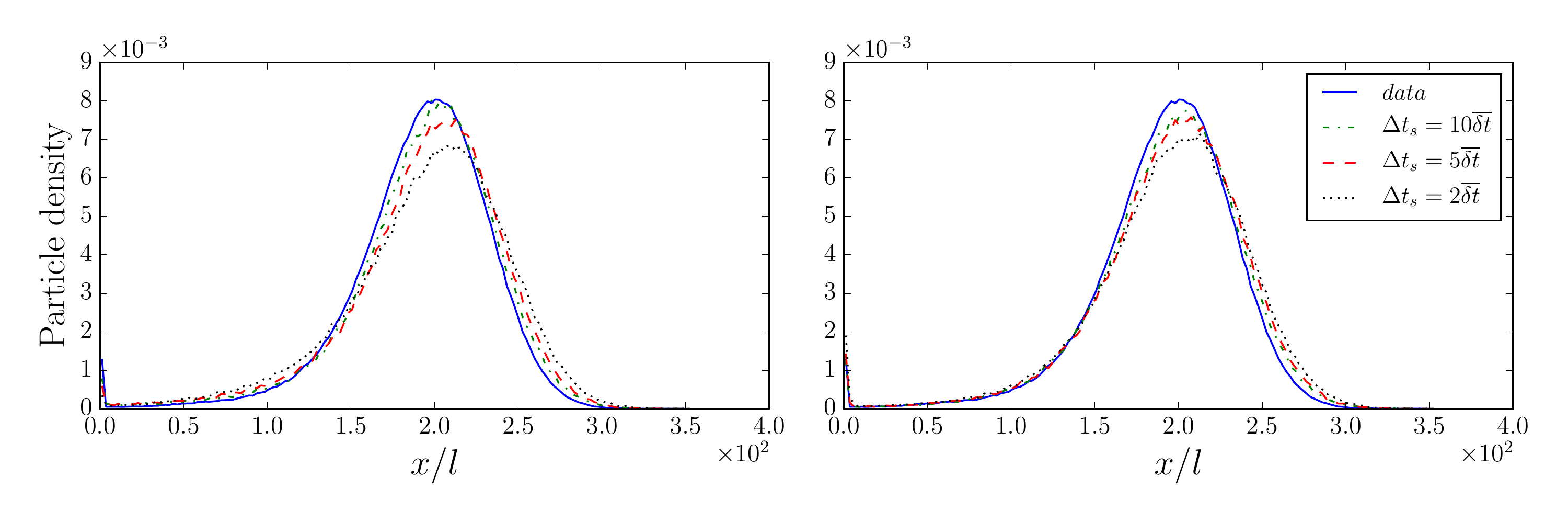}
\caption{Particle density at $t/ \overline{\delta t} = 320 $ for different stencil times; left:
stencil method; right: extended stencil method.} 
\label{dt10_c}
\end{figure*}

Although the extended stencil results in more accurate predictions for $\Delta t_s =
20\overline{\delta t}$, the improvements can be better observed for smaller stencil times.
Figure~\ref{dt10_bt} indicates that there is a significant difference between the predictions of the
stencil model and the extended stencil model for a wide range of first passages times. Most notably,
the slow tail of the FPT curve is captured well with the extended stencil method even for very small
stencil times. Hence, by using the extended stencil we were able to expand the range of accuracy of
temporal Markov models to smaller time steps. In Fig.~\ref{dt10_c} the plumes predicted by the two
stencil models are compared for small stencil times. Although the predictions are not as accurate as
in the case with $\Delta t_s = 20\overline{\delta t}$, for $\Delta t_s = 10\overline{\delta t}$
there still is a good agreement between the model predictions and the MC simulation results. For
smaller stencil times, the extended velocity-angle stencil leads to smaller errors, especially for
the
slow tail of the plume.  \\

\begin{figure*}
\centering
\includegraphics[width=\textwidth]{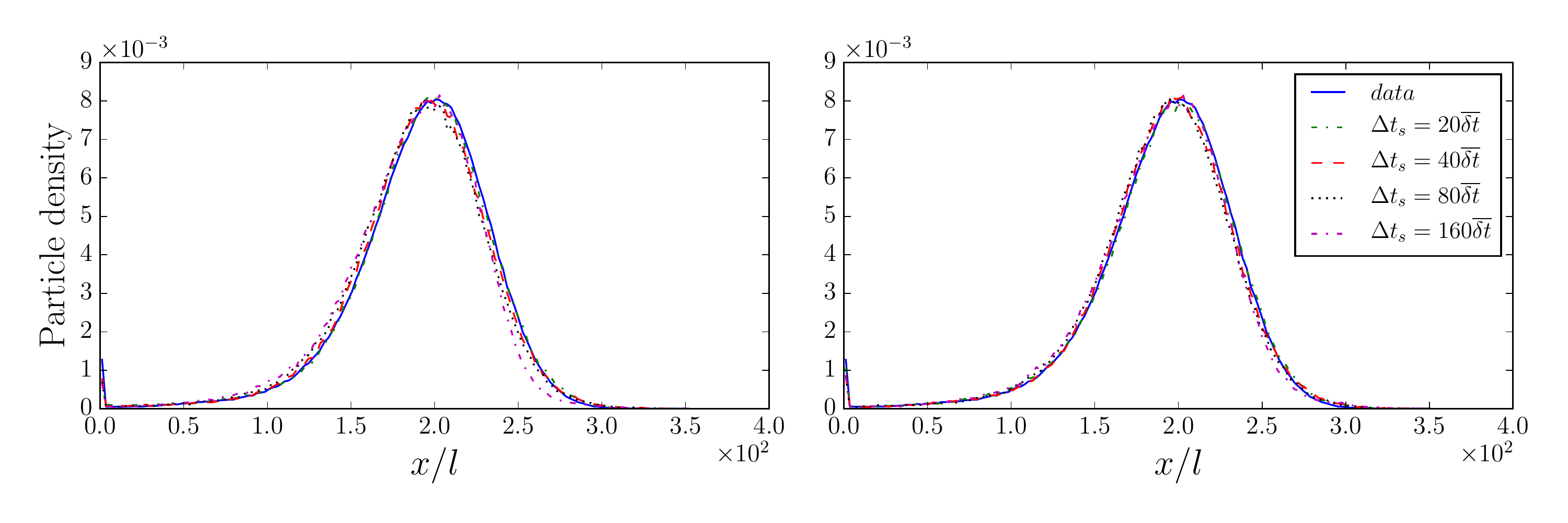}
\caption{Predicted plume concentration for different values of $\Delta t_s / \overline{\delta t}$;
left: stencil method; right: extended stencil method.}
\label{plumes_side_large_dt}
\end{figure*}

\begin{figure*}
\centering
\includegraphics[width=\textwidth]{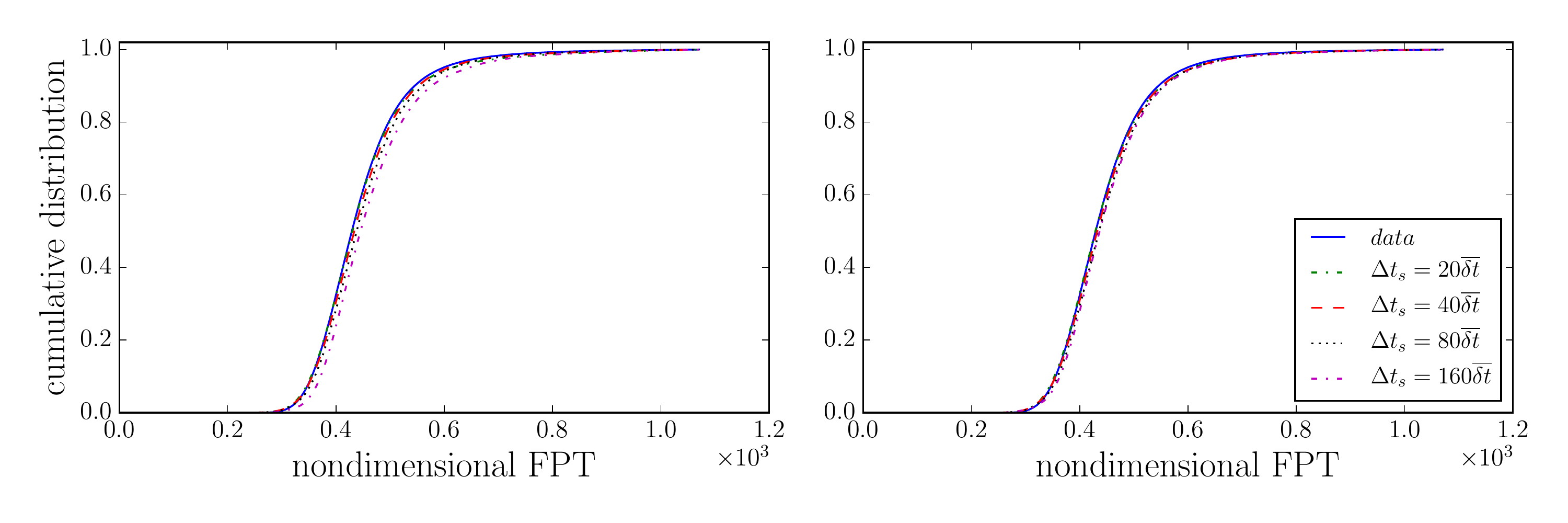}
\caption{FPT curve for different values of $\Delta t_s / \overline{\delta t}$; left: stencil method;
right: extended stencil method.}
\label{cdf_side_large_dt}
\end{figure*}

Given the same MC simulation data, with larger averaging windows there would be fewer velocity
transitions along every streamline and converging to the correct discrete transition matrix would
become harder.  Figures~\ref{plumes_side_large_dt} and~\ref{cdf_side_large_dt} show the predicted
plume and the FPT curve for $\Delta t_s$ equal to $20$, $40$, $80$, $160$ times $\overline{\delta
t}$ for both stencil methods. As can be seen in these figures, both models make similar
predictions and capture the dispersion process well for stencil times up to $\Delta t_s = 80
\overline{\delta t}$. For the largest stencil time ($\Delta t_s = 160 \overline{\delta t}$) the
extended stencil clearly leads to more accurate predictions. 
One should note that for very large stencil times, the MC data is not sufficient to obtain
statistical 
convergence of the transition matrices. This is due to having only a few velocity transitions per
particle trajectory.
However, the extended stencil by construction can capture the extremely slow tail of the plume more 
accurately; this advantage is reflected in the predicted probability densities and FPT curves.\\

\begin{figure*}
\centering
\includegraphics[width=\textwidth]{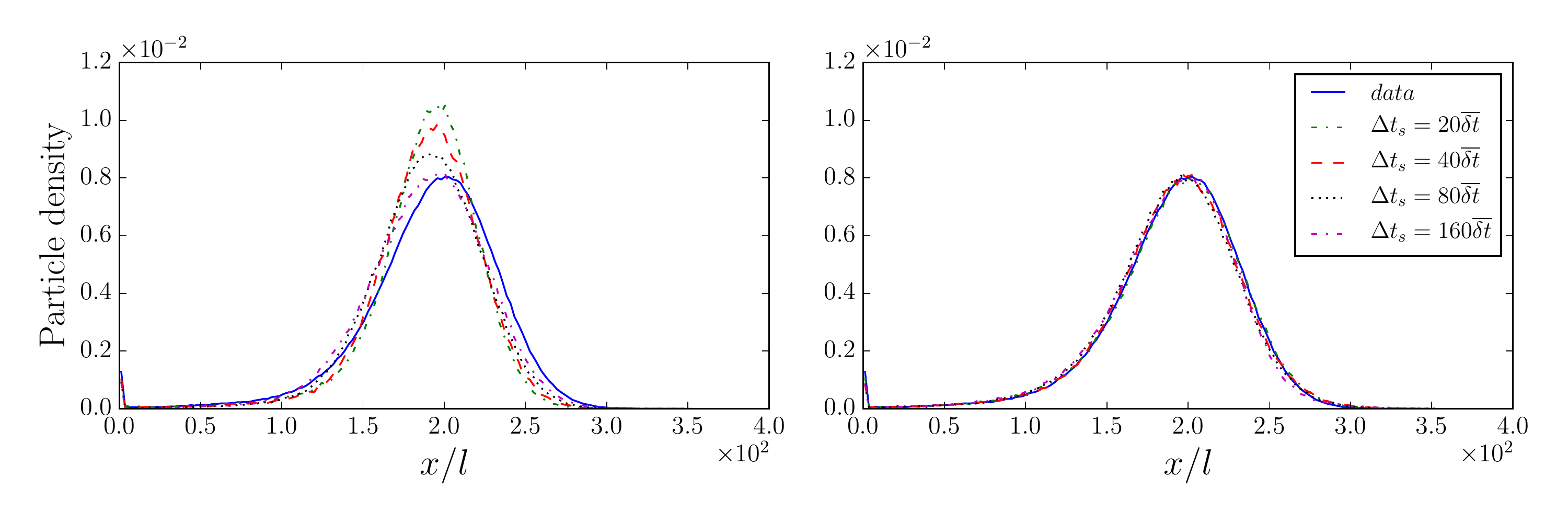}
\caption{Predicted plume concentration for different values of $\Delta t_s / \overline{\delta t}$,
left: using uncorrelated average velocities; right: extended stencil method.}
\label{indPlume}
\end{figure*}

\begin{figure*}
\centering
\includegraphics[width=\textwidth]{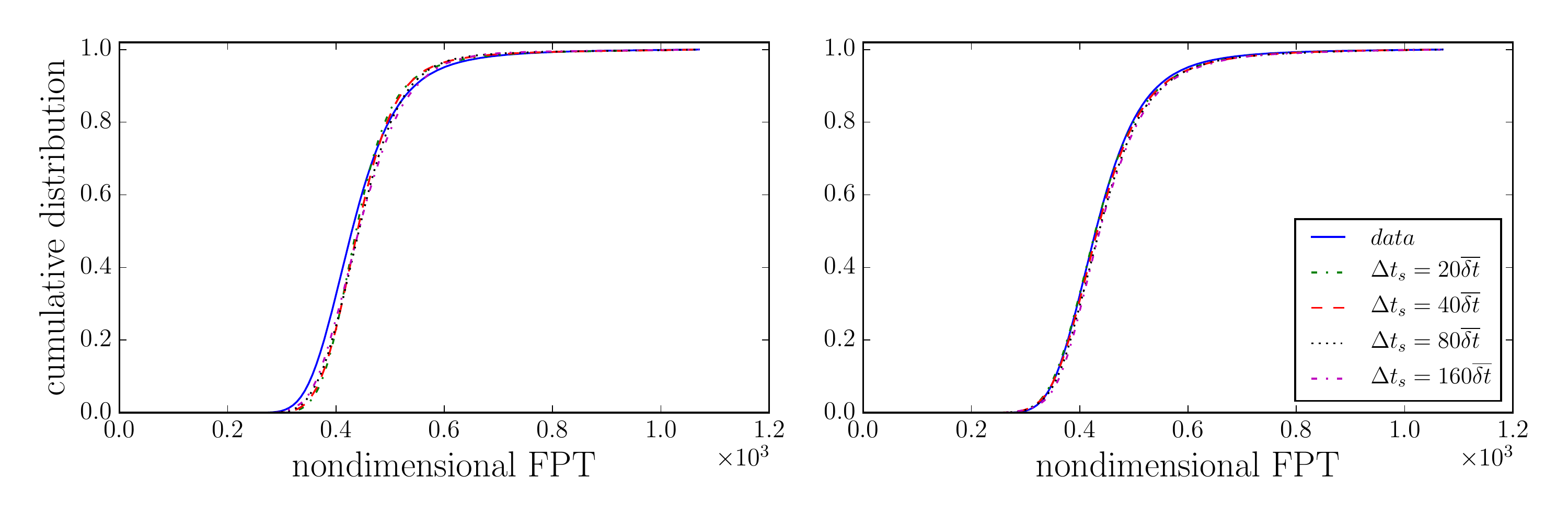}
\caption{FPT curve for different values of $\Delta t_s / \overline{\delta t}$, left: using
uncorrelated average velocities; right: extended stencil method.}
\label{indBT}
\end{figure*}

On the other hand, with larger stencil times the average velocities
would become less correlated. One can argue that with very large stencil times, using a correlated
random walk model would no longer be necessary and independent spatial increments would be
sufficient for modeling the dispersion process at very low temporal resolutions.
Figures~\ref{indPlume} and~\ref{indBT} compare the results of using an uncorrelated temporal random
walk for predicting the FPT and the particle concentration PDF with the results of the extended
stencil model. These results suggest that for a wide range of stencil times ($20 < \Delta t_s /
\overline{\delta t} < 160$) considering the velocity correlation is indeed necessary for making
accurate predictions.\\

The computational cost of the stencil method was compared to correlated CTRW with spatial
transitions that are equal to the link lengths in the network \citep{kang2011spatial}. The cost is
computed by counting the average number of velocity 
transitions along a particle trajectory before exiting the
domain. A stencil model with $\Delta t_s = \alpha \overline{\delta
t}$, with $\alpha > 10$, leads to $\alpha$ times less collisions compared with following every 
transition to a new node;
therefore, it is $\alpha$ times computationally more efficient. For temporal models with $\Delta t_s
<
10\overline{\delta t}$, the stencil method is less accurate than correlated CTRW and does not offer
significant computational gains.\\
 The transition matrices for both the stencil method and the extended stencil method are stored in
 sparse format which results in efficient use of memory. For more details regarding the model
 implementation please refer to the provided repository.


\clearpage
 \section{Application to unstructured networks}
\begin{figure*}
 \centering
 \includegraphics[width=0.4\textwidth]{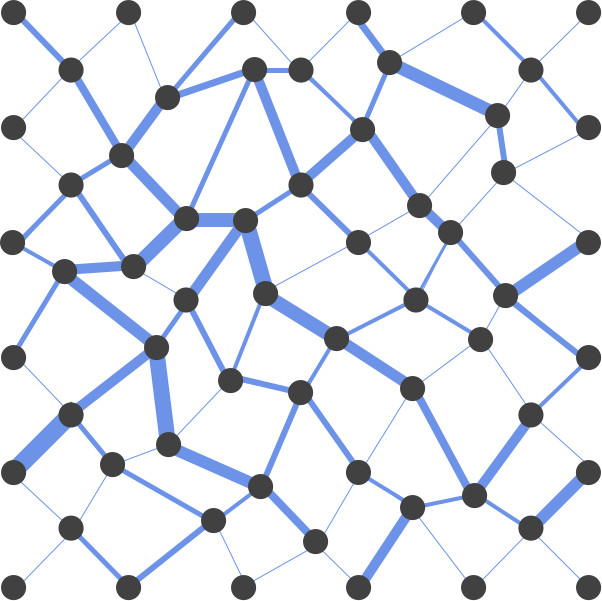}
 \caption{Schematic of an unstructured network generated by adding normal random perturbations to a
 zig zag structured network.}
 \label{unstructured}
 \end{figure*}

\begin{figure*}
 \centering
 \includegraphics[width=\figsize]{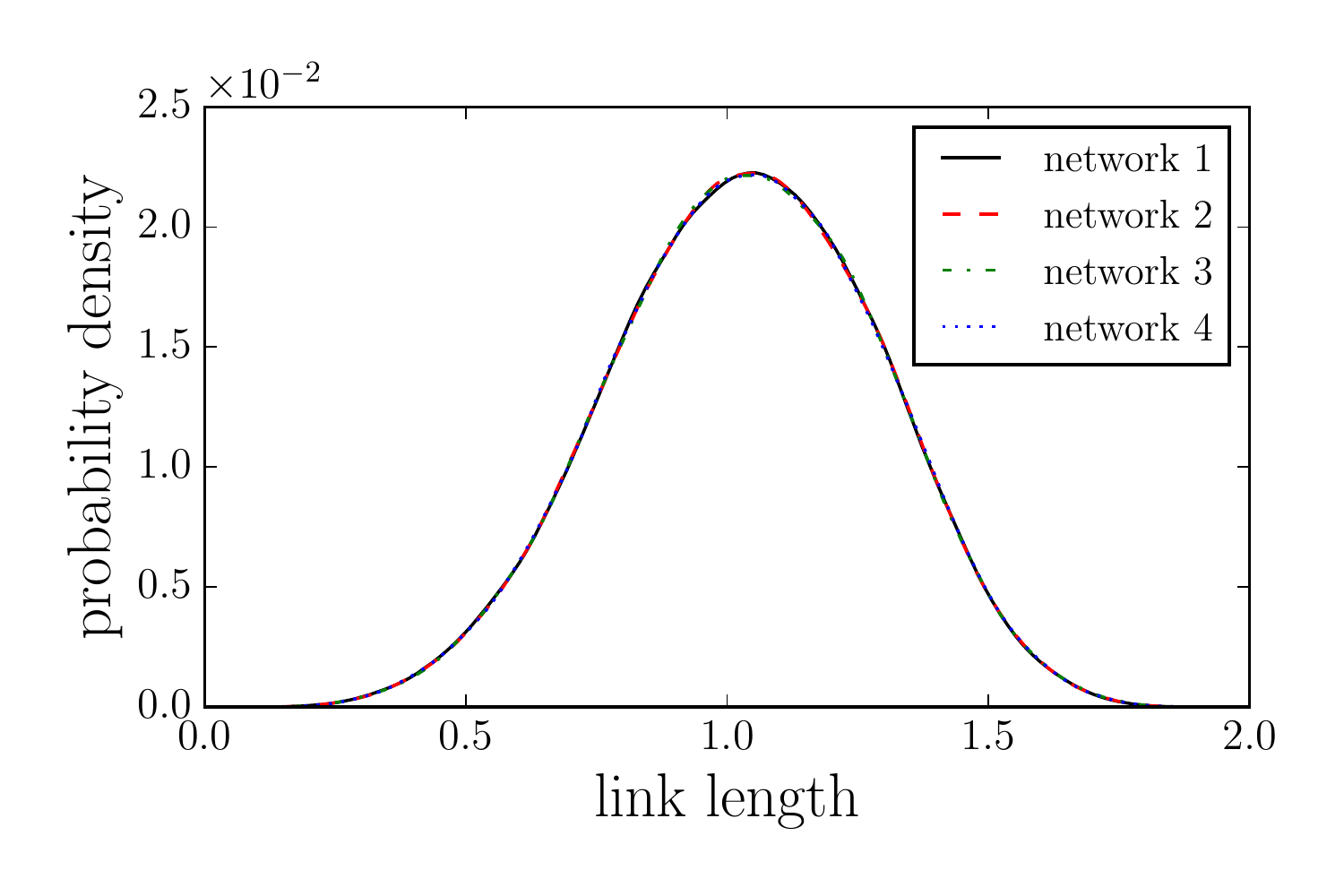}
 \caption{Link length distribution from four realizations of the studied random unstructured
 network.}
 \label{l_distrib}
 \end{figure*}

 The proposed stencil methods require no further generalizations to model transport in unstructured
 networks. In contrast, published correlated spatial models used for simulating transport in
 structured networks are different from the models applied to unstructured networks. In
 \citep{kang2014}, independent spatial Markov processes are used for each spatial dimension, and in
 \citep{kang2017anomalous} the analysis of unstructured fracture networks is limited to the
 projection
 of the particle velocities on the longitudinal direction. Here we illustrate that the proposed
 temporal models are readily applicable to unstructured cases. To this end, unstructured networks
 were generated by randomly perturbing the nodes in the structured network used in the previous
 section.  Normal random perturbations were added independently in the horizontal and vertical
 directions.  A schematic of such a resulting unstructured network is shown in
 Fig.~\ref{unstructured}. This procedure would result in unstructured networks with a truncated
 Gaussian link length distribution.  The link length distribution for four such networks are
 depicted
 in Fig.~\ref{l_distrib}.\\

\begin{figure*}
 \centering
 \includegraphics[width=\textwidth]{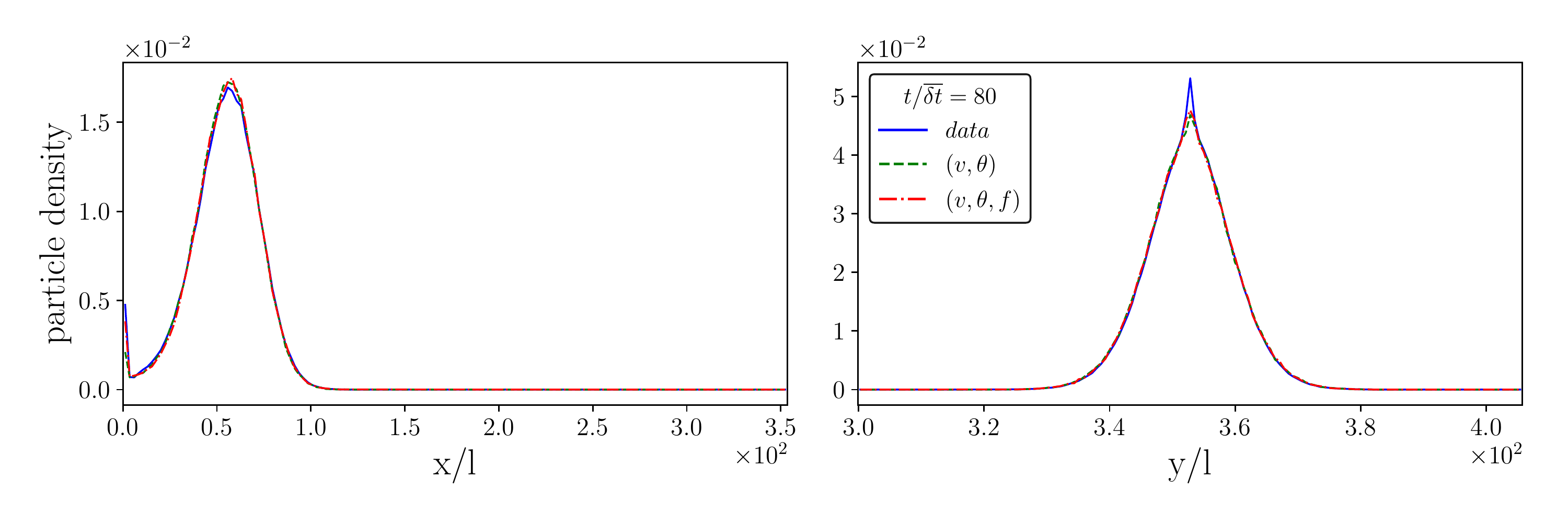}
 \caption{Longitudinal (left) and transverse (right) distributions of the plume at $t/
 \overline{\delta t} = 80$ for
 the unstructured random network test case with $\Delta t_s = 20\overline{\delta t}$.}
 \label{spread_purt1}
 \end{figure*}

\begin{figure*}
 \centering
 \includegraphics[width=\textwidth]{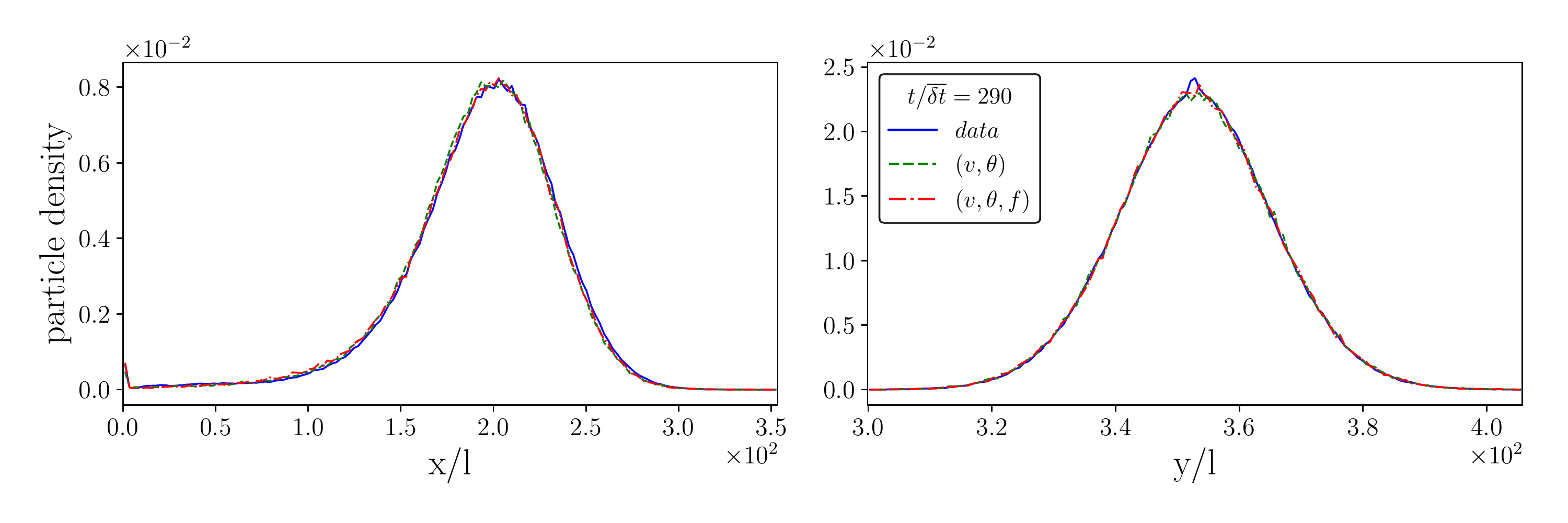}
 \caption{Same as Fig.~\ref{spread_purt1}, at a later time.} 
 \label{spread_purt2}
 \end{figure*}

\begin{figure*}
 \centering
 \includegraphics[width=\figsize]{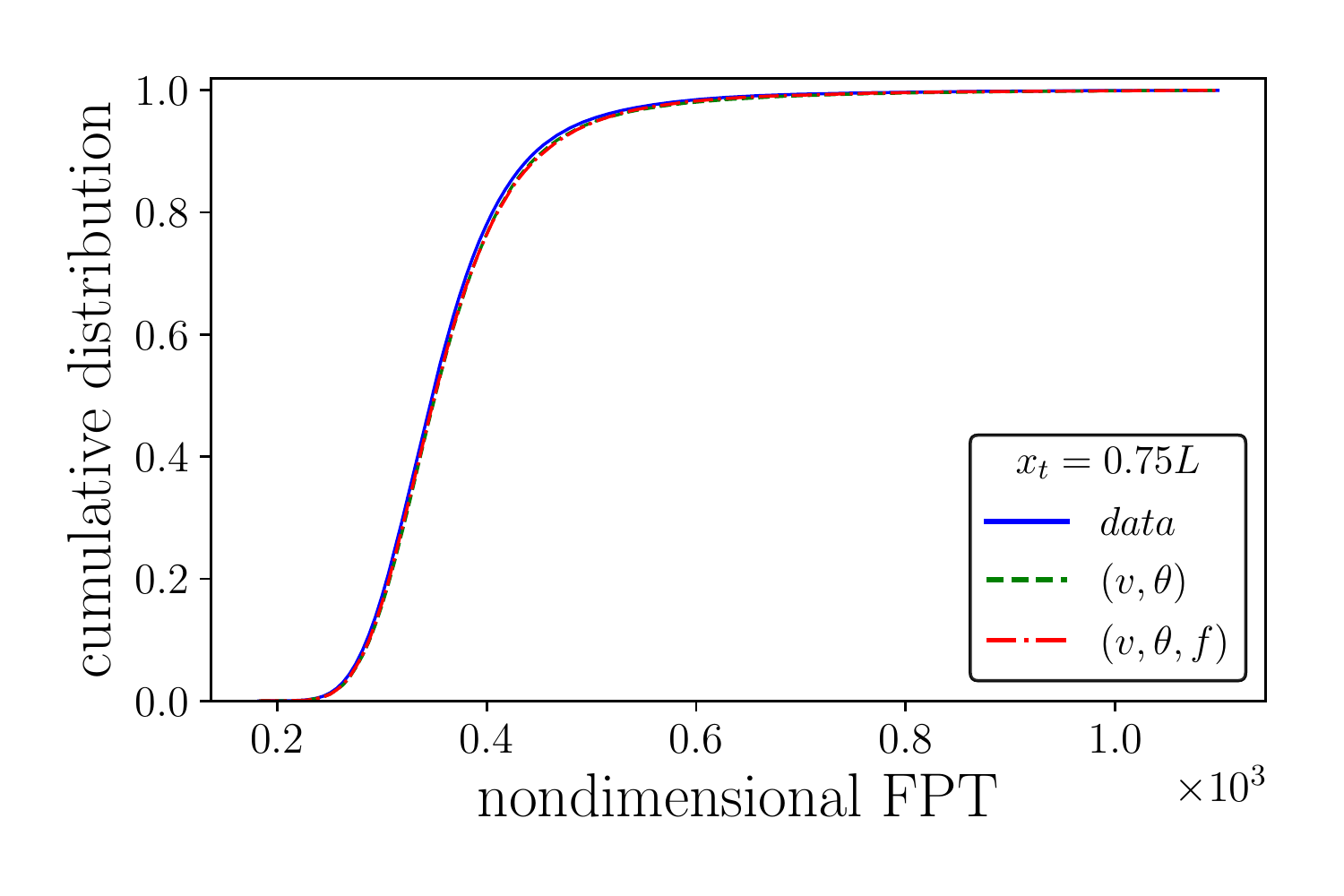}
 \caption{First passage time CDF  for $x_t = 0.75L$ for the unstructured random network example.} 
 \label{bt_p}
 \end{figure*}

 The same particle tracking problem described in section \ref{ptrack} was performed on $1000$
 realizations of these unstructured networks and the problem was also modeled using both the
 velocity-angle and the extended velocity-angle stencils.
 Figures~\ref{spread_purt1} and \ref{spread_purt2} show the plume spreading for two different times. 
 These results show that the stencil model can also be used to accurately predict contaminant
 plume spreading in unstructured networks. Figure~\ref{bt_p} shows that the early and late particle
 arrival times are also captured accurately for this test case.\\

 \section{Conclusions}
 In this work, we showed that temporal Markov models can be used to model transport in random
 networks with good accuracy for a wide range of stencil times.  Although it is known that using
 temporal 
 Markov models can lead to wrong transition rates both from and to low-velocity states, it was shown
 that the
 error induced due to this fact is in many cases small, and can be further reduced by using the
 extended 
 velocity-angle stencil. We were able to improve the results obtained by
 temporal Markov models by enriching the state space of the model to include information on the
 number of repetitions of a given velocity class and extend the range of applicability of these
 models to
 smaller time steps. The extended stencil enhanced the accuracy of predicting the slow
 tail of the contaminant plume, the CDF of first passage time, and the evolution of the second
 central moment of the plume in the longitudinal direction. We also showed that discrete temporal
 Markov
 models can be used to model transport in unstructured networks without any further modification.
 Moreover, since many node transitions can be treated collectively in one stencil step, the proposed
 stencil
 method is more efficient than simulating particle evolution based on following every transition 
 to a new node. Compared to previously proposed temporal Markov models based on SDEs, the models
 proposed
 here make significantly fewer modeling assumptions. 
 Applying the stencil method to correlated heterogeneous fields will be the subject of further
 studies.

\section*{Acknowledgements}
The data used for generating the plots and the code used for generating the complete data set and
the Markov models is available at \url{https://github.com/amirdel/dispersion-random-network}
\citep{delgoshaieMarkovData}. The authors commit to making their data available for at least five
years and refer the reader to the corresponding author.  Amir H. Delgoshaie is grateful to Daniel
Meyer and Oliver Brenner from the Institute of Fluid Dynamics at ETH Zurich for several helpful
discussions and Peter Glynn from the MS\&E department at Stanford University for very helpful
lessons and comments on stochastic modeling. Funding for this project was provided by the Stanford
University Reservoir Simulation Industrial Affiliates (SUPRI--B) program.
\clearpage


\begin{thebibliography}{37}
\providecommand{\natexlab}[1]{#1}
\providecommand{\url}[1]{\texttt{#1}}
\providecommand{\href}[2]{#2}
\providecommand{\path}[1]{#1}
\providecommand{\DOIprefix}{doi:}
\providecommand{\ArXivprefix}{arXiv:}
\providecommand{\URLprefix}{URL: }
\providecommand{\Pubmedprefix}{pmid:}
\providecommand{\doi}[1]{\href{http://dx.doi.org/#1}{\path{#1}}}
\providecommand{\Pubmed}[1]{\href{pmid:#1}{\path{#1}}}
\providecommand{\BIBand}{and}
\providecommand{\bibinfo}[2]{#2}
\ifx\xfnm\undefined \def\xfnm[#1]{\unskip,\space#1}\fi
\makeatletter\def\@biblabel#1{#1.}\makeatother
\bibitem[{Berkowitz et~al.(2006)Berkowitz, Cortis, Dentz and
  Scher}]{berkowitz2006modeling}
\bibinfo{author}{Berkowitz\xfnm[ B.]}, \bibinfo{author}{Cortis\xfnm[ A.]},
  \bibinfo{author}{Dentz\xfnm[ M.]}, \bibinfo{author}{Scher\xfnm[ H.]}.
\newblock \bibinfo{title}{Modeling non-fickian transport in geological
  formations as a continuous time random walk}.
\newblock \emph{\bibinfo{journal}{Reviews of Geophysics}}
  \bibinfo{year}{2006};\bibinfo{volume}{44}(\bibinfo{number}{2}).
\bibitem[{Bouchaud and Georges(1990)}]{bouchaud1990anomalous}
\bibinfo{author}{Bouchaud\xfnm[ J.P.]}, \bibinfo{author}{Georges\xfnm[ A.]}.
\newblock \bibinfo{title}{Anomalous diffusion in disordered media: statistical
  mechanisms, models and physical applications}.
\newblock \emph{\bibinfo{journal}{Physics reports}}
  \bibinfo{year}{1990};\bibinfo{volume}{195}(\bibinfo{number}{4-5}):\bibinfo{pages}{127--293}.
\bibitem[{Edery et~al.(2014)Edery, Guadagnini, Scher and
  Berkowitz}]{edery2014origins}
\bibinfo{author}{Edery\xfnm[ Y.]}, \bibinfo{author}{Guadagnini\xfnm[ A.]},
  \bibinfo{author}{Scher\xfnm[ H.]}, \bibinfo{author}{Berkowitz\xfnm[ B.]}.
\newblock \bibinfo{title}{Origins of anomalous transport in heterogeneous
  media: Structural and dynamic controls}.
\newblock \emph{\bibinfo{journal}{Water Resources Research}}
  \bibinfo{year}{2014};\bibinfo{volume}{50}(\bibinfo{number}{2}):\bibinfo{pages}{1490--1505}.
\bibitem[{Nowak et~al.(2012)Nowak, Rubin and Barros}]{nowak2012hypothesis}
\bibinfo{author}{Nowak\xfnm[ W.]}, \bibinfo{author}{Rubin\xfnm[ Y.]},
  \bibinfo{author}{Barros\xfnm[ F.P.]}.
\newblock \bibinfo{title}{A hypothesis-driven approach to optimize field
  campaigns}.
\newblock \emph{\bibinfo{journal}{Water Resources Research}}
  \bibinfo{year}{2012};\bibinfo{volume}{48}(\bibinfo{number}{6}).
\bibitem[{Moslehi and de~Barros(2017)}]{moslehi2017uncertainty}
\bibinfo{author}{Moslehi\xfnm[ M.]}, \bibinfo{author}{de~Barros\xfnm[ F.P.]}.
\newblock \bibinfo{title}{Uncertainty quantification of environmental
  performance metrics in heterogeneous aquifers with long-range correlations}.
\newblock \emph{\bibinfo{journal}{Journal of Contaminant Hydrology}}
  \bibinfo{year}{2017};\bibinfo{volume}{196}:\bibinfo{pages}{21--29}.
\bibitem[{Ghorbanidehno et~al.(2015)Ghorbanidehno, Kokkinaki, Li, Darve and
  Kitanidis}]{ghorbanidehno2015real}
\bibinfo{author}{Ghorbanidehno\xfnm[ H.]}, \bibinfo{author}{Kokkinaki\xfnm[
  A.]}, \bibinfo{author}{Li\xfnm[ J.Y.]}, \bibinfo{author}{Darve\xfnm[ E.]},
  \bibinfo{author}{Kitanidis\xfnm[ P.K.]}.
\newblock \bibinfo{title}{Real-time data assimilation for large-scale systems:
  The spectral kalman filter}.
\newblock \emph{\bibinfo{journal}{Advances in Water Resources}}
  \bibinfo{year}{2015};\bibinfo{volume}{86}:\bibinfo{pages}{260--272}.
\bibitem[{Li et~al.(2015)Li, Kokkinaki, Ghorbanidehno, Darve and
  Kitanidis}]{li2015compressed}
\bibinfo{author}{Li\xfnm[ J.Y.]}, \bibinfo{author}{Kokkinaki\xfnm[ A.]},
  \bibinfo{author}{Ghorbanidehno\xfnm[ H.]}, \bibinfo{author}{Darve\xfnm[
  E.F.]}, \bibinfo{author}{Kitanidis\xfnm[ P.K.]}.
\newblock \bibinfo{title}{The compressed state kalman filter for nonlinear
  state estimation: Application to large-scale reservoir monitoring}.
\newblock \emph{\bibinfo{journal}{Water Resources Research}}
  \bibinfo{year}{2015};.
\bibitem[{Ghorbanidehno et~al.(2017)Ghorbanidehno, Kokkinaki, Kitanidis and
  Darve}]{ghorbanidehno2017optimal}
\bibinfo{author}{Ghorbanidehno\xfnm[ H.]}, \bibinfo{author}{Kokkinaki\xfnm[
  A.]}, \bibinfo{author}{Kitanidis\xfnm[ P.K.]}, \bibinfo{author}{Darve\xfnm[
  E.]}.
\newblock \bibinfo{title}{Optimal estimation and scheduling in aquifer
  management using the rapid feedback control method}.
\newblock \emph{\bibinfo{journal}{Advances in Water Resources}}
  \bibinfo{year}{2017};\bibinfo{volume}{110}:\bibinfo{pages}{310--318}.
\bibitem[{Fiori et~al.(2015)Fiori, Zarlenga, Gotovac, Jankovic, Volpi,
  Cvetkovic and Dagan}]{fiori2015advective}
\bibinfo{author}{Fiori\xfnm[ A.]}, \bibinfo{author}{Zarlenga\xfnm[ A.]},
  \bibinfo{author}{Gotovac\xfnm[ H.]}, \bibinfo{author}{Jankovic\xfnm[ I.]},
  \bibinfo{author}{Volpi\xfnm[ E.]}, \bibinfo{author}{Cvetkovic\xfnm[ V.]},
  \bibinfo{author}{Dagan\xfnm[ G.]}.
\newblock \bibinfo{title}{Advective transport in heterogeneous aquifers: Are
  proxy models predictive?}
\newblock \emph{\bibinfo{journal}{Water resources research}}
  \bibinfo{year}{2015};\bibinfo{volume}{51}(\bibinfo{number}{12}):\bibinfo{pages}{9577--9594}.
\bibitem[{Banton et~al.(1997)Banton, Delay and Porel}]{banton1997new}
\bibinfo{author}{Banton\xfnm[ O.]}, \bibinfo{author}{Delay\xfnm[ F.]},
  \bibinfo{author}{Porel\xfnm[ G.]}.
\newblock \bibinfo{title}{A new time domain random walk method for solute
  transport in 1--d heterogeneous media}.
\newblock \emph{\bibinfo{journal}{Ground Water}}
  \bibinfo{year}{1997};\bibinfo{volume}{35}(\bibinfo{number}{6}):\bibinfo{pages}{1008--1013}.
\bibitem[{Bodin et~al.(2003)Bodin, Porel and Delay}]{bodin2003simulation}
\bibinfo{author}{Bodin\xfnm[ J.]}, \bibinfo{author}{Porel\xfnm[ G.]},
  \bibinfo{author}{Delay\xfnm[ F.]}.
\newblock \bibinfo{title}{Simulation of solute transport in discrete fracture
  networks using the time domain random walk method}.
\newblock \emph{\bibinfo{journal}{Earth and Planetary Science Letters}}
  \bibinfo{year}{2003};\bibinfo{volume}{208}(\bibinfo{number}{3}):\bibinfo{pages}{297--304}.
\bibitem[{Graham and McLaughlin(1989)}]{WRCR:WRCR4722}
\bibinfo{author}{Graham\xfnm[ W.]}, \bibinfo{author}{McLaughlin\xfnm[ D.]}.
\newblock \bibinfo{title}{Stochastic analysis of nonstationary subsurface
  solute transport: 1. unconditional moments}.
\newblock \emph{\bibinfo{journal}{Water Resources Research}}
  \bibinfo{year}{1989};\bibinfo{volume}{25}(\bibinfo{number}{2}):\bibinfo{pages}{215--232}.
\newblock \URLprefix \url{http://dx.doi.org/10.1029/WR025i002p00215}.
  \DOIprefix\doi{10.1029/WR025i002p00215}.
\bibitem[{Le~Borgne et~al.(2007)Le~Borgne, de~Dreuzy, Davy and
  Bour}]{le2007characterization}
\bibinfo{author}{Le~Borgne\xfnm[ T.]}, \bibinfo{author}{de~Dreuzy\xfnm[ J.R.]},
  \bibinfo{author}{Davy\xfnm[ P.]}, \bibinfo{author}{Bour\xfnm[ O.]}.
\newblock \bibinfo{title}{Characterization of the velocity field organization
  in heterogeneous media by conditional correlation}.
\newblock \emph{\bibinfo{journal}{Water resources research}}
  \bibinfo{year}{2007};\bibinfo{volume}{43}(\bibinfo{number}{2}).
\bibitem[{Le~Borgne et~al.(2008{\natexlab{a}})Le~Borgne, Dentz and
  Carrera}]{le2008lagrangian}
\bibinfo{author}{Le~Borgne\xfnm[ T.]}, \bibinfo{author}{Dentz\xfnm[ M.]},
  \bibinfo{author}{Carrera\xfnm[ J.]}.
\newblock \bibinfo{title}{Lagrangian statistical model for transport in highly
  heterogeneous velocity fields}.
\newblock \emph{\bibinfo{journal}{Physical review letters}}
  \bibinfo{year}{2008}{\natexlab{a}};\bibinfo{volume}{101}(\bibinfo{number}{9}):\bibinfo{pages}{090601}.
\bibitem[{Le~Borgne et~al.(2008{\natexlab{b}})Le~Borgne, Dentz and
  Carrera}]{le2008spatial}
\bibinfo{author}{Le~Borgne\xfnm[ T.]}, \bibinfo{author}{Dentz\xfnm[ M.]},
  \bibinfo{author}{Carrera\xfnm[ J.]}.
\newblock \bibinfo{title}{Spatial markov processes for modeling lagrangian
  particle dynamics in heterogeneous porous media}.
\newblock \emph{\bibinfo{journal}{Physical Review E}}
  \bibinfo{year}{2008}{\natexlab{b}};\bibinfo{volume}{78}(\bibinfo{number}{2}):\bibinfo{pages}{026308}.
\bibitem[{Kang et~al.(2011)Kang, Dentz, Le~Borgne and Juanes}]{kang2011spatial}
\bibinfo{author}{Kang\xfnm[ P.K.]}, \bibinfo{author}{Dentz\xfnm[ M.]},
  \bibinfo{author}{Le~Borgne\xfnm[ T.]}, \bibinfo{author}{Juanes\xfnm[ R.]}.
\newblock \bibinfo{title}{Spatial markov model of anomalous transport through
  random lattice networks}.
\newblock \emph{\bibinfo{journal}{Physical review letters}}
  \bibinfo{year}{2011};\bibinfo{volume}{107}(\bibinfo{number}{18}):\bibinfo{pages}{180602}.
\bibitem[{Kang et~al.(2015{\natexlab{a}})Kang, Dentz, Le~Borgne and
  Juanes}]{kang2015anomalous}
\bibinfo{author}{Kang\xfnm[ P.K.]}, \bibinfo{author}{Dentz\xfnm[ M.]},
  \bibinfo{author}{Le~Borgne\xfnm[ T.]}, \bibinfo{author}{Juanes\xfnm[ R.]}.
\newblock \bibinfo{title}{Anomalous transport on regular fracture networks:
  Impact of conductivity heterogeneity and mixing at fracture intersections}.
\newblock \emph{\bibinfo{journal}{Physical Review E}}
  \bibinfo{year}{2015}{\natexlab{a}};\bibinfo{volume}{92}(\bibinfo{number}{2}):\bibinfo{pages}{022148}.
\bibitem[{Kang et~al.(2015{\natexlab{b}})Kang, Le~Borgne, Dentz, Bour and
  Juanes}]{kang2015impact}
\bibinfo{author}{Kang\xfnm[ P.K.]}, \bibinfo{author}{Le~Borgne\xfnm[ T.]},
  \bibinfo{author}{Dentz\xfnm[ M.]}, \bibinfo{author}{Bour\xfnm[ O.]},
  \bibinfo{author}{Juanes\xfnm[ R.]}.
\newblock \bibinfo{title}{Impact of velocity correlation and distribution on
  transport in fractured media: Field evidence and theoretical model}.
\newblock \emph{\bibinfo{journal}{Water Resources Research}}
  \bibinfo{year}{2015}{\natexlab{b}};\bibinfo{volume}{51}(\bibinfo{number}{2}):\bibinfo{pages}{940--959}.
\bibitem[{Kang et~al.(2014)Kang, {De Anna}, Nunes, Bijeljic, Blunt and
  Juanes}]{kang2014}
\bibinfo{author}{Kang\xfnm[ P.K.]}, \bibinfo{author}{{De Anna}\xfnm[ P.]},
  \bibinfo{author}{Nunes\xfnm[ J.P.]}, \bibinfo{author}{Bijeljic\xfnm[ B.]},
  \bibinfo{author}{Blunt\xfnm[ M.J.]}, \bibinfo{author}{Juanes\xfnm[ R.]}.
\newblock \bibinfo{title}{{Pore-Scale intermittent velocity structure
  underpinning anomalous transport through 3-D porousmedia}}.
\newblock \emph{\bibinfo{journal}{Geophysical Research Letters}}
  \bibinfo{year}{2014};\bibinfo{volume}{41}(\bibinfo{number}{17}):\bibinfo{pages}{6184--6190}.
\newblock \DOIprefix\doi{10.1002/2014GL061475}.
\bibitem[{Kang et~al.(2017)Kang, Dentz, Le~Borgne, Lee and
  Juanes}]{kang2017anomalous}
\bibinfo{author}{Kang\xfnm[ P.K.]}, \bibinfo{author}{Dentz\xfnm[ M.]},
  \bibinfo{author}{Le~Borgne\xfnm[ T.]}, \bibinfo{author}{Lee\xfnm[ S.]},
  \bibinfo{author}{Juanes\xfnm[ R.]}.
\newblock \bibinfo{title}{Anomalous transport in disordered fracture networks:
  spatial markov model for dispersion with variable injection modes}.
\newblock \emph{\bibinfo{journal}{Advances in Water Resources}}
  \bibinfo{year}{2017};.
\bibitem[{Meyer and Tchelepi(2010)}]{meyer2010particle}
\bibinfo{author}{Meyer\xfnm[ D.W.]}, \bibinfo{author}{Tchelepi\xfnm[ H.A.]}.
\newblock \bibinfo{title}{Particle-based transport model with markovian
  velocity processes for tracer dispersion in highly heterogeneous porous
  media}.
\newblock \emph{\bibinfo{journal}{Water Resources Research}}
  \bibinfo{year}{2010};\bibinfo{volume}{46}(\bibinfo{number}{11}).
\bibitem[{Meyer et~al.(2010)Meyer, Jenny and Tchelepi}]{meyer2010joint}
\bibinfo{author}{Meyer\xfnm[ D.W.]}, \bibinfo{author}{Jenny\xfnm[ P.]},
  \bibinfo{author}{Tchelepi\xfnm[ H.A.]}.
\newblock \bibinfo{title}{A joint velocity-concentration pdf method for tracer
  flow in heterogeneous porous media}.
\newblock \emph{\bibinfo{journal}{Water Resources Research}}
  \bibinfo{year}{2010};\bibinfo{volume}{46}(\bibinfo{number}{12}).
\bibitem[{Meyer and Saggini(2016)}]{meyer2016testing}
\bibinfo{author}{Meyer\xfnm[ D.W.]}, \bibinfo{author}{Saggini\xfnm[ F.]}.
\newblock \bibinfo{title}{Testing the markov hypothesis in fluid flows}.
\newblock \emph{\bibinfo{journal}{Physical Review E}}
  \bibinfo{year}{2016};\bibinfo{volume}{93}(\bibinfo{number}{5}):\bibinfo{pages}{053103}.
\bibitem[{Meyer et~al.(2013)Meyer, Tchelepi and Jenny}]{meyer2013fast}
\bibinfo{author}{Meyer\xfnm[ D.W.]}, \bibinfo{author}{Tchelepi\xfnm[ H.A.]},
  \bibinfo{author}{Jenny\xfnm[ P.]}.
\newblock \bibinfo{title}{A fast simulation method for uncertainty
  quantification of subsurface flow and transport}.
\newblock \emph{\bibinfo{journal}{Water Resources Research}}
  \bibinfo{year}{2013};\bibinfo{volume}{49}(\bibinfo{number}{5}):\bibinfo{pages}{2359--2379}.
\bibitem[{Meyer and Bijeljic(2016)}]{meyer2016pore}
\bibinfo{author}{Meyer\xfnm[ D.W.]}, \bibinfo{author}{Bijeljic\xfnm[ B.]}.
\newblock \bibinfo{title}{Pore-scale dispersion: Bridging the gap between
  microscopic pore structure and the emerging macroscopic transport behavior}.
\newblock \emph{\bibinfo{journal}{Physical Review E}}
  \bibinfo{year}{2016};\bibinfo{volume}{94}(\bibinfo{number}{1}):\bibinfo{pages}{013107}.
\bibitem[{Attinger et~al.(2004)Attinger, Dentz and
  Kinzelbach}]{attinger2004exact}
\bibinfo{author}{Attinger\xfnm[ S.]}, \bibinfo{author}{Dentz\xfnm[ M.]},
  \bibinfo{author}{Kinzelbach\xfnm[ W.]}.
\newblock \bibinfo{title}{Exact transverse macro dispersion coefficients for
  transport in heterogeneous porous media}.
\newblock \emph{\bibinfo{journal}{Stochastic Environmental Research and Risk
  Assessment}}
  \bibinfo{year}{2004};\bibinfo{volume}{18}(\bibinfo{number}{1}):\bibinfo{pages}{9--15}.
\bibitem[{Blunt et~al.(2002)Blunt, Jackson, Piri and
  Valvatne}]{blunt2002detailed}
\bibinfo{author}{Blunt\xfnm[ M.J.]}, \bibinfo{author}{Jackson\xfnm[ M.D.]},
  \bibinfo{author}{Piri\xfnm[ M.]}, \bibinfo{author}{Valvatne\xfnm[ P.H.]}.
\newblock \bibinfo{title}{Detailed physics, predictive capabilities and
  macroscopic consequences for pore-network models of multiphase flow}.
\newblock \emph{\bibinfo{journal}{Advances in Water Resources}}
  \bibinfo{year}{2002};\bibinfo{volume}{25}(\bibinfo{number}{8}):\bibinfo{pages}{1069--1089}.
\bibitem[{Dong and Blunt(2009)}]{dong2009pore}
\bibinfo{author}{Dong\xfnm[ H.]}, \bibinfo{author}{Blunt\xfnm[ M.J.]}.
\newblock \bibinfo{title}{Pore-network extraction from
  micro-computerized-tomography images}.
\newblock \emph{\bibinfo{journal}{Physical review E}}
  \bibinfo{year}{2009};\bibinfo{volume}{80}(\bibinfo{number}{3}):\bibinfo{pages}{036307}.
\bibitem[{Khayrat and Jenny(2016)}]{khayrat2016subphase}
\bibinfo{author}{Khayrat\xfnm[ K.]}, \bibinfo{author}{Jenny\xfnm[ P.]}.
\newblock \bibinfo{title}{Subphase approach to model hysteretic two-phase flow
  in porous media}.
\newblock \emph{\bibinfo{journal}{Transport in Porous Media}}
  \bibinfo{year}{2016};\bibinfo{volume}{111}(\bibinfo{number}{1}):\bibinfo{pages}{1--25}.
\bibitem[{Mehmani and Tchelepi(2017)}]{mehmani2017minimum}
\bibinfo{author}{Mehmani\xfnm[ Y.]}, \bibinfo{author}{Tchelepi\xfnm[ H.A.]}.
\newblock \bibinfo{title}{Minimum requirements for predictive pore-network
  modeling of solute transport in micromodels}.
\newblock \emph{\bibinfo{journal}{Advances in Water Resources}}
  \bibinfo{year}{2017};.
\bibitem[{Delgoshaie et~al.(2015)Delgoshaie, Meyer, Jenny and
  Tchelepi}]{delgoshaie2015non}
\bibinfo{author}{Delgoshaie\xfnm[ A.H.]}, \bibinfo{author}{Meyer\xfnm[ D.W.]},
  \bibinfo{author}{Jenny\xfnm[ P.]}, \bibinfo{author}{Tchelepi\xfnm[ H.A.]}.
\newblock \bibinfo{title}{Non-local formulation for multiscale flow in porous
  media}.
\newblock \emph{\bibinfo{journal}{Journal of Hydrology}}
  \bibinfo{year}{2015};\bibinfo{volume}{531}:\bibinfo{pages}{649--654}.
\bibitem[{Jenny and Meyer(2016)}]{jenny2016non}
\bibinfo{author}{Jenny\xfnm[ P.]}, \bibinfo{author}{Meyer\xfnm[ D.]}.
\newblock \bibinfo{title}{Non-local generalization of darcy’s law based on
  empirically extracted conductivity kernels}.
\newblock In: \emph{\bibinfo{booktitle}{ECMOR XV-15th European Conference on
  the Mathematics of Oil Recovery}}. \bibinfo{year}{2016}:\unskip.
\bibitem[{Kang et~al.(2016)Kang, Brown and Juanes}]{kang2016emergence}
\bibinfo{author}{Kang\xfnm[ P.K.]}, \bibinfo{author}{Brown\xfnm[ S.]},
  \bibinfo{author}{Juanes\xfnm[ R.]}.
\newblock \bibinfo{title}{Emergence of anomalous transport in stressed rough
  fractures}.
\newblock \emph{\bibinfo{journal}{Earth and Planetary Science Letters}}
  \bibinfo{year}{2016};\bibinfo{volume}{454}:\bibinfo{pages}{46--54}.
\bibitem[{Painter and Cvetkovic(2005)}]{painter2005upscaling}
\bibinfo{author}{Painter\xfnm[ S.]}, \bibinfo{author}{Cvetkovic\xfnm[ V.]}.
\newblock \bibinfo{title}{Upscaling discrete fracture network simulations: An
  alternative to continuum transport models}.
\newblock \emph{\bibinfo{journal}{Water Resources Research}}
  \bibinfo{year}{2005};\bibinfo{volume}{41}(\bibinfo{number}{2}).
\bibitem[{D{\"u}nser and Meyer(2016)}]{dunser2016predicting}
\bibinfo{author}{D{\"u}nser\xfnm[ S.]}, \bibinfo{author}{Meyer\xfnm[ D.W.]}.
\newblock \bibinfo{title}{Predicting field-scale dispersion under realistic
  conditions with the polar markovian velocity process model}.
\newblock \emph{\bibinfo{journal}{Advances in Water Resources}}
  \bibinfo{year}{2016};\bibinfo{volume}{92}:\bibinfo{pages}{271--283}.
\bibitem[{Resnick(2013)}]{resnick2013adventures}
\bibinfo{author}{Resnick\xfnm[ S.I.]}.
\newblock \bibinfo{title}{Adventures in stochastic processes}.
\newblock \bibinfo{publisher}{Springer Science \& Business Media};
  \bibinfo{year}{2013}.
\bibitem[{Delgoshaie(2018)}]{delgoshaieMarkovData}
\bibinfo{author}{Delgoshaie\xfnm[ A.H.]}.
\newblock \bibinfo{title}{amirdel/dispersion-random-network: random network}.
\newblock \bibinfo{year}{2018}.
\newblock \URLprefix \url{https://doi.org/10.5281/zenodo.1213380}.
  \DOIprefix\doi{10.5281/zenodo.1213380}.

\end{thebibliography}

\end{document}